\documentclass{tlp}
\usepackage[utf8]{inputenc}
\usepackage{amsmath}
\usepackage{graphicx}
\usepackage{multirow}

\usepackage{xcolor}
\usepackage{xspace}
\usepackage{comment}
\usepackage{tikz}
\usepackage{amsfonts,amsmath,amsfonts,amssymb}
\usepackage[style=base]{caption}
\usepackage[style=base]{subcaption}
\usepackage{algorithm}
\usepackage[noend]{algpseudocode}
\usepackage{tabularx}
\usepackage{booktabs}
\usepackage[xcolor]{changebar}
\newtheorem{example}{Example}[section]

\newcommand{\diamondplus}{%
  \raisebox{-.15ex}{\begin{tikzpicture}
    \useasboundingbox (-0.8ex, -0.8ex) rectangle (0.8ex, 0.8ex);
    \node (w) at (-0.8ex,0) {};
    \node (e) at (+0.8ex,0) {};
    \node (s) at (0,-.8ex) {};
    \node (n) at (0,+.8ex) {};
    \draw [-] (n.center) -- (e.center) -- (s.center) -- (w.center) -- (n.center);
    \draw [-] (n.center) -- (s.center);
    \draw [-] (e.center) -- (w.center);
  \end{tikzpicture}}}
\newcommand{\diamondminus}{%
  \raisebox{-.15ex}{\begin{tikzpicture}
    \useasboundingbox (-0.8ex, -0.8ex) rectangle (0.8ex, 0.8ex);
    \node (w) at (-0.8ex,0) {};
    \node (e) at (+0.8ex,0) {};
    \node (s) at (0,-.8ex) {};
    \node (n) at (0,+.8ex) {};
    \draw [-] (n.center) -- (e.center) -- (s.center) -- (w.center) -- (n.center);
    \draw [-] (e.center) -- (w.center);
  \end{tikzpicture}}}    
  
\newcommand{\SinceOperator}{\mathbin{\mathcal{S}}}
\newcommand{\UntilOperator}{\mathbin{\mathcal{U}}}

\newcommand{\LiviaComments}[1]{\textcolor{purple}{ }}

\newcommand{\Revision}[1]{#1}

\newcommand{\RevisionEnv}[1]{#1}
\newcommand{\RevisionEnvTwo}[1]{#1}
\newcommand{\RevisionEnvThree}[1]{#1}

\newcommand{\Remove}[1]{\textcolor{red}{}}
\newcommand{\RevisionTwo}[1]{#1}

\newcommand{\RevisionThree}[1]{#1}

\usepackage{mdframed}

\begin{document}

\lefttitle{Bellomarini, L., Blasi, L., Nissl, M., Sallinger, E.}

\jnlPage{1}{25}
\jnlDoiYr{2021}
\doival{10.1017/xxxxx}

\title[Theory and Practice of Logic Programming]{The Temporal Vadalog System:\\ Temporal Datalog-based Reasoning}

\begin{authgrp}
\author{\gn{Luigi} \sn{Bellomarini} }
\affiliation{Bank of Italy, Italy}
\author{\gn{Livia} \sn{Blasi} }
\affiliation{Bank of Italy, Italy} \affiliation{TU Wien, Austria}
\author{\gn{Markus} \sn{Nissl}}
\affiliation{TU Wien, Austria}
\author{\gn{Emanuel} \sn{Sallinger}}
\affiliation{TU Wien, Austria} \affiliation{University of Oxford, UK}
\end{authgrp}

\history{\sub{31-03-2023}}

\maketitle

\begin{abstract}
In the wake of the recent resurgence of the Datalog language of databases, together with its extensions for ontological reasoning settings, this work aims to bridge the gap between the theoretical studies of DatalogMTL (Datalog extended with metric temporal logic) and the development of production-ready reasoning systems. In particular, we lay out the functional and architectural desiderata of a modern reasoner and propose our system, Temporal Vadalog. Leveraging the vast amount of experience from the database community, we go beyond the typical chase-based implementations of reasoners, and propose a set of novel techniques and a system that adopts a modern data pipeline architecture. We discuss crucial architectural choices, such as how to guarantee termination when infinitely many time intervals are possibly generated, how to merge intervals, and how to sustain a limited memory footprint. We discuss advanced features of the system, such as the support for time series, and present an extensive experimental evaluation. 

This paper is a substantially extended version of ``The Temporal Vadalog System'' as presented at RuleML+RR '22. 

Under consideration in Theory and Practice of Logic Programming (TPLP).
\end{abstract}


\section{Introduction}
In recent years, the Datalog language~\citep{datalog2} has been experiencing renewed interest, thanks to the growing adoption for \emph{Knowledge Representation and Reasoning} applications~\citep{DBLP:conf/aiia/Gottlob22}. The additional requirements introduced by reasoning spawned prolific research towards new extensions of the language to support advanced features such as existential quantification and aggregation. These extensions are frequently shared under the common name of Datalog$^\pm$~\citep{DBLP:journals/ws/CaliGL12} and include languages, technically, ``fragments'', that exhibit a very good trade-off between computational complexity and expressive power, such as Warded and Shy Datalog$^\pm$~\citep{DBLP:conf/kr/BaldazziBFS22}.

\smallskip
As the new logical languages flourish,
the expectations for declarative, explainable 
expressivity
of complex domains, jointly with
feasible computation, are rekindling the idea of using
deductive languages in practical applications, with modern systems~\citep{LAAC19} based on Datalog in production settings, often in combination with knowledge-based AI models, such as \textit{knowledge graphs}~\citep{BFGS19}.  

\smallskip
With
so many real-world adoptions, the temporal perspective is increasingly becoming a first-class requirement of Datalog. In fact, with the pervasive constant and voluminous streams of data,
we are witnessing that temporal reasoning---the ability to reason about events and intervals in time---is becoming of the essence in many fields, from healthcare to finance, robotics to transportation. We can look at a plethora of use cases: the longitudinal studies of anticancer treatments to explain and prevent adverse events~\citep{MaLeeTemporalHealth21}, the analysis and prediction of stock market time series~\citep{WaKC19}, the tracking of objects for unmanned 
vehicles~\citep{LinTemporalRoboticsTransportation}, the processing of IoT devices interactions~\citep{WCKK19}, and the exploitation of log data~\citep{BrKR18}.

\smallskip
The 
proposal for an extension of Datalog with \textit{Metric Temporal Logic} operators, or DatalogMTL~\citep{BrKR18}, shows an extremely promising impulse towards a declarative language
with
sufficient expressive power to handle reasoning settings, such as recursion,
while offering time awareness within a computationally effective framework.

\smallskip
\noindent\textbf{Positioning of the paper}.
This paper studies the theoretical and practical underpinnings at the foundation of modern implementations of DatalogMTL. We illustrate the functional and architectural desiderata of such systems and present the \textit{Temporal Vadalog} system, a fully-engineered reasoning DatalogMTL tool. 
\Revision{This work is a substantially extended version of a RuleML+RR '22 paper~\citep{BBNS22} introducing the system.}
Besides proposing a broader and completely reformulated presentation of the content, the contribution is here enriched with (i) an in-depth discussion about the logical optimization of the temporal operators; (ii) the implementation of new temporal operators (namely, \textit{since} and \textit{until}); (iii) an entirely new part dedicated to the implementation of time series in Temporal Vadalog; (iv) a corpus of new experiments. 

\smallskip
To
present the ideas at the foundation of Temporal Vadalog, let us start introducing DatalogMTL with 
a 
financial use case
example
from our work with the Bank of Italy.

\begin{mdframed}
\begin{example}
\label{ex:running_example} 
%\textit{
Changes in strategically important companies may face government scrutiny to uphold regulatory standards, focusing on both the companies and their new shareholders.
We describe the scenario 
by a database $\mathcal{D}$ of facts about company shares and the following set $\Pi$ of DatalogMTL rules.
{\small
\begin{align*}
    \boxplus_{[0,1]} \mathit{significantShare}(X,Y),  \\
    \neg \diamondminus_{[0,1]} \mathit{significantShare}(X,Y) & \to \mathit{significantOwner}(X,Y) \tag{1}\\
    \mathit{watchCompany}(Y), \mathit{significantOwner}(X,Y),  \\
    \mathit{connected}(X,Z) & \to \mathit{watchCompany}(Z) \tag{2}
\end{align*}}
\noindent
We assume the reader is familiar with 
Datalog rule syntax:
%the Datalog syntax of rules: 
the left-hand side of the rule (the body) is the implication premise and is composed of a logical conjunction of predicates over terms, i.e., constants or variables; the right-hand side (the head) is the implication conclusion. 
In the rules, we find predicates from the relational schema of $\mathcal{D}$. In particular, \emph{significantShare} holds whenever $X$ retains a high number of shares of $Y$; \emph{significantOwner} is then a derived, namely, ``\emph{intensional}'' predicate, obtained by Rule~1. Then, by Rule~2, whenever $Y$ is interested in a \emph{watchCompany} fact to indicate that the authority deems that company of some strategic relevance, given that $X$ is a significant owner of $Y$, then all companies $Z$ 
%that are 
\emph{connected} to $X$ will be watched over as well. Neglecting for a moment the temporal angle of $\Pi$, its semantics is straightforward: whenever the premise of one rule holds with respect to $\mathcal{D}$, then $\mathcal{D}$ is augmented with new facts to satisfy the implication if it does not already hold. For instance, if $\mathcal{D}$ contains the facts $\mathit{watchCompany}(\mathit{ACME})$, $\textit{significantOwner}(A,\mathit{ACME})$, $\mathit{connected}(A,\mathit{EMCA})$, then the new fact $\textit{watchCompany}(\mathit{EMCA})$ will be added to $\mathcal{D}$. Here, the temporal perspective is introduced by the $\boxplus$ and $\diamondminus$ operators. When prefixed with 
$\boxplus$,
%the $\boxplus$ operator, 
a predicate holds, only if the predicate itself continuously holds in the interval ($[t,t+1]$, if Rule~1 is evaluated at $t$) in the future, as indicated by the operator internal in the index. The $\diamondminus$ operator modifies a predicate's semantics so that it holds if the predicate itself holds at least once in the past interval (\Revision{$[t-1,t]$,} if Rule~1 is evaluated at $t$). Clearly, in a temporal interpretation of Datalog, all facts, including the ``\emph{extensional}'' ones of $\mathcal{D}$, are time-annotated. In the example, Rule~1 fires only if $X$, who has never been a significant shareholder for $Y$ in the past, becomes such in the future.
\end{example}
\end{mdframed}

\smallskip
\noindent
More formally speaking, the semantics of a set of Datalog rules
%(or program) 
$\Pi$ is defined via the \textsc{chase} procedure~\citep{MaMS79}, which modifies $\mathcal{D}$ as long as all the rules of $\Pi$ are satisfied. Multiple variants of the \Revision{\textsc{chase}} arose, including many %that support 
supporting
DatalogMTL. 
At the same time, several DatalogMTL fragments were also introduced, to carefully balance expressive power and computational complexity:
%. Among the recent ones, for example, 
DatalogMTL$^{\text{FP}}$~\citep{WCKK19} and its core and linear eponymous DatalogMTL$_{\textit{core}}^{\diamondminus}$ and DatalogMTL$_{\textit{lin}}^{\diamondminus}$~\citep{DBLP:conf/ijcai/WalegaGKK20}, or Integer DatalogMTL~\citep{DBLP:conf/kr/WalegaGKK20} offer a good prospect for reasoning over time, \Revision{with some of them being tractable, i.e., \textit{query answering} can be evaluated in polynomial time with respect to the database size.}
%On the other hand, 
Moreover,
the system should support operations over numeric values and aggregations, as well as negation. 

\smallskip
To get to the core of our discussion, let us lay out the characteristics that a reasoning system should support to deal with time as a first-class system, via DatalogMTL.

\smallskip\noindent\textbf{Functional Desiderata.} A temporal reasoning system is a ``reasoner'' in the first place. A good yardstick for the features it should incorporate then comes from the experience with \textit{knowledge graph management systems}: such tools should adopt a simple, modular, low-complexity, and highly expressive language. \Revision{DatalogMTL fragments like the ones mentioned above should then be supported.} To cope with real-world applications, monotonic aggregations should be supported as well~\citep{DBLP:conf/ruleml/BellomariniNS21}.

\smallskip\noindent\textbf{Architectural Desiderata.} The existing systems 
implementing DatalogMTL adopt time-aware extensions of the {\sc chase}. Yet, the vast 
experience from the database community suggests that a direct implementation of the \Revision{\textsc{chase}} can be a suboptimal choice, affected by multiple limitations. 
For efficiency and memory footprint, the extensional database and all intermediate facts must be in memory. We wish for reasoning systems to be memory-bound, using a modern data processing pipeline architecture.
As for desiderata specific to temporal reasoners,
temporal programs can produce facts that periodically hold over time, with repeating patterns potentially generating infinite facts, which practically can be represented compactly depending on the DatalogMTL fragment. A temporal reasoner should recognize these fragments, identify the patterns at runtime, and guarantee termination. 
Architecturally, \RevisionThree{it} should have a configurable toolbox of strategies for different fragments.
Time intervals should be carefully handled, with a selection of algorithms to merge and compress them for a space-efficient representation.

\smallskip
The development of temporal reasoners is in its infancy, with the only proposals for practical implementation of DatalogMTL being either not engineered for production use or not supporting essential features such as recursion, like in the system by~\cite{AAAI1714881}. Conversely, MeTeoR~\citep{DBLP:conf/aaai/WangHWG22} supports recursion; still, it does not offer operations over numeric values or aggregations. 

\smallskip\noindent\textbf{Contribution.} 
In this paper, we present the Temporal Vadalog system, a fully-engineered reasoning system that addresses the above desiderata.
Functionally, it offers:

\begin{itemize}
\item support for \textbf{multiple DatalogMTL fragments}, including DatalogMTL$^{FP}$ (\Revision{i.e., \textit{forward-propagating};} resp. BP\Revision{, i.e., \textit{backward-propagating})} \RevisionTwo{over the real timeline, }and DatalogMTL$^{FP}$ \Revision{(resp. BP)} over the integer timeline
   \item native support of the \textbf{DatalogMTL operators} (box, diamond, since and until)
        \item operations over \textbf{numeric} values
        \item  \textbf{recursion}, \textbf{aggregations}, and \textbf{stratified negation}
        \item \textbf{time series} and time series operators.
\end{itemize}

\noindent
Its architecture is based on a modern data pipeline style, where programs are evaluated in a \textit{pull-based} and \textit{query-driven} manner, sustaining limited memory footprint and high scalability for medium- to high-size settings. More in particular: 

\begin{itemize}
    \item \textbf{rewriting-based optimization}
    \item fragment-aware \textbf{termination control strategies}
    \item seamless integration of \textbf{temporal and non-temporal reasoning}.
\end{itemize}

\smallskip
\noindent\textbf{Overview.} Section~\ref{sec:preliminaries} presents \Revision{DatalogMTL
%, its syntax, 
and an overview of time series operators.} Section~\ref{sec:temporal-vadalog-architecture} covers the main contribution of the paper, detailing the Temporal Vadalog pipeline and its components; Section~\ref{sec:time-series} shows 
how the system can be used to reason
over time series. Section~\ref{sec:experiments} presents the empirical evaluation of the system with a discussion of the experiments. In Section~\ref{sec:related-work} we discuss the related work; finally, we conclude the paper in Section~\ref{sec:conclusions}. Additional material can be found in the Appendix~\citep{appendix}.

\section{Preliminaries}
\label{sec:preliminaries}

\subsection{DatalogMTL}
\label{sec:datalogmtl} 
DatalogMTL is Datalog extended with operators from the metric temporal logic. 
This section is a summary of DatalogMTL with stratified negation under continuous semantics.

DatalogMTL is defined over the rational timeline, i.e., an ordered set of rational numbers $\mathbb{Q}$. An interval $\varrho = \langle \varrho^-, \varrho^+ \rangle $ is a non-empty subset of $\mathbb{Q}$ such that for each $t \in \mathbb{Q}$ where $\varrho^- < t < \varrho^+$, $t \in \varrho$, and the endpoints $\varrho^-, \varrho^+ \in \mathbb{Q} \cup \{-\infty, \infty \}$. The brackets denote whether the interval is closed (``$[]$''), half-open (``$[)$'',``$(]$'') or open (``$()$'')
\Revision{, whereas angle brackets (``$\langle \rangle$'') are used when unspecified.}
An interval is \emph{punctual} if it is of the form $[t,t]$, 
%or abbreviated just $t$,
\emph{positive} if $\varrho^- \geq 0$, and \emph{bounded} if $\varrho^-, \varrho^+ \in \mathbb{Q}$.

\smallskip
DatalogMTL extends the syntax of Datalog with negation with temporal operators~\citep{DBLP:conf/aaai/CucalaWGK21}. 
For the following definitions, we consider a function-free first-order signature. 
An \emph{atom} is of the form $P(\boldsymbol{\tau})$, where $P$ is a $n$-ary predicate and $\boldsymbol{\tau}$ is a $n$-ary tuple of terms, where a term is either a constant or a variable. 
An atom is \emph{ground} if it contains no variables. 
A \emph{fact} is an expression $P(\boldsymbol{\tau})@\varrho$, where $\varrho$ is an interval and $P(\boldsymbol{\tau})$ a ground atom and a \emph{database} is a set of facts.
A \emph{literal} is an expression given by the following grammar, where $\varrho$ is a positive interval:
$A ::=  \top \mid \bot \mid P(\boldsymbol{\tau}) \mid \boxminus_\varrho A \mid \boxplus_\varrho A \mid 
    \diamondminus_\varrho A \mid \diamondplus_\varrho A \mid A \SinceOperator_\varrho A \mid A \UntilOperator_\varrho A$.
A \emph{rule} is an expression given by the following grammar, where $i,j \geq 0$, each $A_k$ ($k \geq 0$) is a literal and $B$ is an atom: 
$A_1 \land \dots \land A_i \land \mathrm{not}\;  A_{i+1} \land \dots \land \mathrm{not}\; A_{i+j} \to B$.
\Revision{The conjunction of literals $A_k$ is the rule body, where $A_1 \land \dots \land A_i$ denote positive literals and $A_{i+1} \land \dots \land A_{i+j}$ denote negated (i.e., prefixed with \textit{not}) literals.} The atom $B$ is the rule head.
A rule is \emph{safe} if each variable occurs in at least one positive body literal, \emph{positive} if it has no negated body literals (i.e., $j=0$), and \emph{ground} if it contains no variables.
A program $\Pi$ is a set of safe rules and is \emph{stratifiable} if there exists a stratification of a program $\Pi$. 
A stratification of $\Pi$ is defined as a function $\sigma$ that maps each predicate $P$ in $\Pi$ to a positive integer \RevisionThree{(\textit{stratum})} s.t.\ for each rule, where $P^{h}$ denotes a predicate of the head, and $P^{+}$ (resp. $P^{-}$) a positive (negative) body predicate, $\sigma(P^{h}) \geq \sigma(P^+)$ and $\sigma(P^{h}) > \sigma(P^-)$.

\smallskip
The semantics of DatalogMTL is given by an interpretation $\mathfrak{M}$ that specifies for each time point $t \in \mathbb{Q}$ and each ground atom $P(\boldsymbol{\tau})$, whether $P(\boldsymbol{\tau})$ is satisfied at $t$, in which case we write $\mathfrak{M}, t \models P(\boldsymbol{\tau})$. This satisfiability notion extends to ground literals as follows:
\begin{align*}
\footnotesize
\begin{aligned}
    & \mathfrak{M}, t \models \top & \textrm{ for each } t\\
    & \mathfrak{M}, t \models \bot & \textrm{ for no } t\\
    & \mathfrak{M}, t \models \boxminus_\varrho A & \textrm{iff } \mathfrak{M}, s \models A \textrm{ for all } s \textrm{ with } t-s \in \varrho \\
    & \mathfrak{M}, t \models \boxplus_\varrho A & \textrm{iff } \mathfrak{M}, s \models A \textrm{ for all } s \textrm{ with } s-t \in \varrho \\
    & \mathfrak{M}, t \models A \SinceOperator_\varrho A' & \textrm{iff } \mathfrak{M}, s \models A' \textrm{ for some } s \textrm{ with } t-s \in \varrho  \land~\mathfrak{M}, r \models A \textrm{ for all } r \in (s,t) \\
    & \mathfrak{M}, t \models A \UntilOperator_\varrho A' & \textrm{iff } \mathfrak{M}, s \models A' \textrm{ for some } s \textrm{ with } s-t \in \varrho  \land~\mathfrak{M}, r \models A \textrm{ for all } r \in (t,s)  \\
    & \mathfrak{M}, t \models \diamondminus_\varrho A & \textrm{iff } \mathfrak{M}, s \models A \textrm{ for some } s \textrm{ with } t-s \in \varrho \\
    & \mathfrak{M}, t \models \diamondplus_\varrho A & \textrm{iff } \mathfrak{M}, s \models A \textrm{ for some } s \textrm{ with } s-t \in \varrho
\end{aligned}
\end{align*}
An interpretation $\mathfrak{M}$ satisfies $\mathrm{not}~A$ ($\mathfrak{M}, t \models \mathrm{not}~A$), if $\mathfrak{M},t \not \models A$, a fact $P(\boldsymbol{\tau})@\varrho$, if $\mathfrak{M}, t \models P(\boldsymbol{\tau})$ for all $t \in \varrho$, and a set of facts $\mathcal{D}$ if it is a model of each fact in $\mathcal{D}$.
Furthermore, $\mathfrak{M}$ satisfies a ground rule $r$ if $\mathfrak{M},t \models A_k$ for  $0 \leq k \leq i$ and $\mathfrak{M},t \models \mathrm{not}~A_k$ for $i+1 \leq k \leq i+j$ for every $t$\RevisionThree{; for every $t$, if the literals in the body are satisfied, so is the head} \RevisionTwo{$\mathfrak{M},t \models B$; $\mathfrak{M}$ satisfies a rule} when it satisfies every possible grounding of the rule. 
Moreover, $\mathfrak{M}$ is a \textit{model} of a program if it satisfies every rule in the program and the program has a stratification, i.e., it is \textit{stratifiable}.
\RevisionTwo{Given a stratifiable program $\Pi$ and a set of facts $\mathcal{D}$, we call $\mathfrak{C}_{\Pi, \mathcal{D}}$  the \textit{canonical model} of $\Pi$ and $\mathcal{D}$~\citep{BrKR18}},\RevisionThree{ and define it as the minimum model of $\Pi$ and $\mathcal{D}$, i.e., $\mathfrak{C}_{\Pi, \mathcal{D}}$ is the minimum model for all the facts of $\mathcal{D}$ and the rules of $\Pi$. In this context, ``minimum'' means that the set of positive literals in $\mathfrak{M}$ is minimized or, equivalently, that the positive literals of this model are contained in every other model. Since $\Pi$ is stratifiable, this minimum model exists and is unique~\citep{DBLP:journals/jacm/GelderRS91}}.
\RevisionTwo{According to Tena Cucala's notation~\citep{DBLP:conf/aaai/CucalaWGK21}, we say that a stratifiable\RevisionTwo{ program $\Pi$} and a set of facts $\mathcal{D}$ entail a fact $P(\boldsymbol{\tau})@\varrho$, written as $(\Pi,\mathcal{D}) \models P(\boldsymbol{\tau})@\varrho$, if $\mathfrak{C}_{\Pi, \mathcal{D}} \models P(\boldsymbol{\tau})@\varrho$. In the remainder of the paper, we will assume the stratification of programs (or set of rules) as implicit.} 

\Revision{In this context, the \textit{query answering} or \textit{reasoning} task is defined as follows:} \RevisionTwo{given the pair $Q = (\Pi, \mathit{Ans})$, where $\Pi$ is a set of rules, $\mathit{Ans}$ is an $n$-ary predicate, and the query $Q$ is evaluated over $\mathcal{D}$, then $Q(\mathcal{D})$ is defined as} \RevisionThree{$Q(\mathcal{D}) = \{(\bar{t}, \varrho) \in \mathit{dom}(\mathcal{D})^{n} \times \mathit{time}(\mathcal{D}) \mid (\Pi, \mathcal{D}) \models \mathit{Ans}(\bar{t})@\varrho\}$}\RevisionTwo{, where $\bar{t}$ is a tuple of terms, the domain of $\mathcal{D}$, denoted $\mathit{dom}(\mathcal{D})$, is the set of all constants that appear in the facts of $\mathcal{D}$, and the set of all the time intervals in $\mathcal{D}$ is denoted as $\mathit{time}(\mathcal{D})$.}
\Revision{As we shall see in practical cases, the $\mathit{Ans}$ predicate of $\Pi$ will be sometimes called ``query predicate'' and provided to the reasoning system with specific conventions, which we omit for space reasons, but will render in textual explanations.}

\smallskip\noindent\Revision{\textbf{Fragments of DatalogMTL.} We define DatalogMTL$^{\text{FP}}$~\citep{WCKK19} as the DatalogMTL language restricted to only make use of \textit{forward-propagating} operators, i.e., the operators that only examine the past, as in $\diamondminus$, $\boxminus$ and $\SinceOperator$. \RevisionTwo{DatalogMTL$^{\text{BP}}$ is symmetrical to DatalogMTL$^{\text{FP}}$ using backward-propagating operators, e.g., $\diamondplus$, $\boxplus$ and $\UntilOperator$.}
A \textit{linear} language has at most one intensional atom in the body of the rules, while in a \textit{core} language, in addition, rules without $\bot$ in the head contain only one body atom. 
DatalogMTL$_{\textit{core}}^{\diamondminus}$, DatalogMTL$_{\textit{lin}}^{\diamondminus}$~\citep{DBLP:conf/ijcai/WalegaGKK20} are thus defined as the \textit{core} and \textit{linear} fragments using only the $\diamondminus$ operator. Finally, Integer DatalogMTL~\citep{DBLP:conf/kr/WalegaGKK20} is DatalogMTL operating exclusively on an integer (discrete) timeline.}

\subsection{Time Series}
\Revision{Time series data is a sequence of data points, collected over time, that measure the performance of a particular variable. For example, we may have time series for stock prices or inflation expectations in finance and economics, humidity readings in domotics, and all the relevant vital measurements for healthcare. }

Time series are typically made of sequences of evenly-spaced punctual temporal intervals, in which case they are considered \textit{regular}; otherwise, they are called \textit{irregular} time series.
The distance between intervals defines the frequency (or time resolution).

To understand how to work with time series, we need a clear overview of the operators that can be applied and how they manipulate and transform data to detect and measure patterns and trends.
Let us go through an introduction to the common time series operators, and we will later explore their implementation in Temporal Vadalog.  

\smallskip\noindent\textbf{Shifting.} The shifting operator moves the time index of the data, allowing to compare data points from different periods (e.g., the same month in different years).

\smallskip\noindent\textbf{Rolling.} The rolling operator calculates a moving time window (rolling window), typically to apply statistical metrics to smooth out short-term data fluctuations.

\smallskip\noindent \textbf{Resampling.} The resampling operator transforms the time series by changing its frequency, by lowering it (\textit{downsampling)} or by increasing it (\textit{upsampling}). 

\smallskip\noindent \textbf{Moving Averages.} 
Moving averages use rolling windows to smooth data and filter noise and fluctuations. Types depend on the different weight over data points, and include: a) \textit{Simple Moving Average}, where all points have equal weight, and the \textit{Centered Moving Average} variant; b) \textit{Exponential Moving Average}, where recent points have more weight.

\smallskip\noindent \textbf{Stock and flow.} Measurements in time series can be stocks, quantities existing at a specific time, or flows, variables measured over an interval of time, comparable to the rate or speed, and can be derived from stocks (and vice versa). For instance, in real estate, the stock could refer to the 
available housing units,
%housing units available in the market, 
while the flow represents the rate of new units entering it.  
For the conversion between measurement units, \RevisionThree{we use} the following operators: a) \textit{Stock to flow}, \Revision{the difference between stocks over consecutive time points;} b) \textit{Flow to stock}, the cumulative sum of a series of flows over time.

\smallskip\noindent\textbf{Seasonal decomposition.} The seasonal decomposition operator splits a time series into its trend, seasonal, and residual components, which helps identify patterns and trends.

\section{The Temporal Vadalog Architecture}
\label{sec:temporal-vadalog-architecture}

\Revision{The architecture aims to support query answering with DatalogMTL: 
given a database of time-annotated facts and a set of rules with one or more query predicates, we produce an execution plan for that query.}
Our approach is inspired by our extensive experience in constructing database and knowledge graph management systems. 
\Revision{Our architecture builds on that of the Vadalog system~\citep{BeSG18}, which in its current version does not support temporal reasoning. Vadalog adopts the \textit{volcano iterator model}~\citep{GrMc93}, thus employing consolidated query evaluation techniques from a reasoning perspective. In this work, we go beyond and extend the approach in such a way that the query plans include temporal operators (i.e., they are \textit{time-aware}).}
Rather than providing a comprehensive taxonomy of all the components in the architecture, we will take a thematic walk-through, with a focus on addressing the challenges of temporal reasoning. In the next section, we will introduce our time-aware execution pipeline.

\begin{figure}
    \centering
    \includegraphics[width=0.68\textwidth]{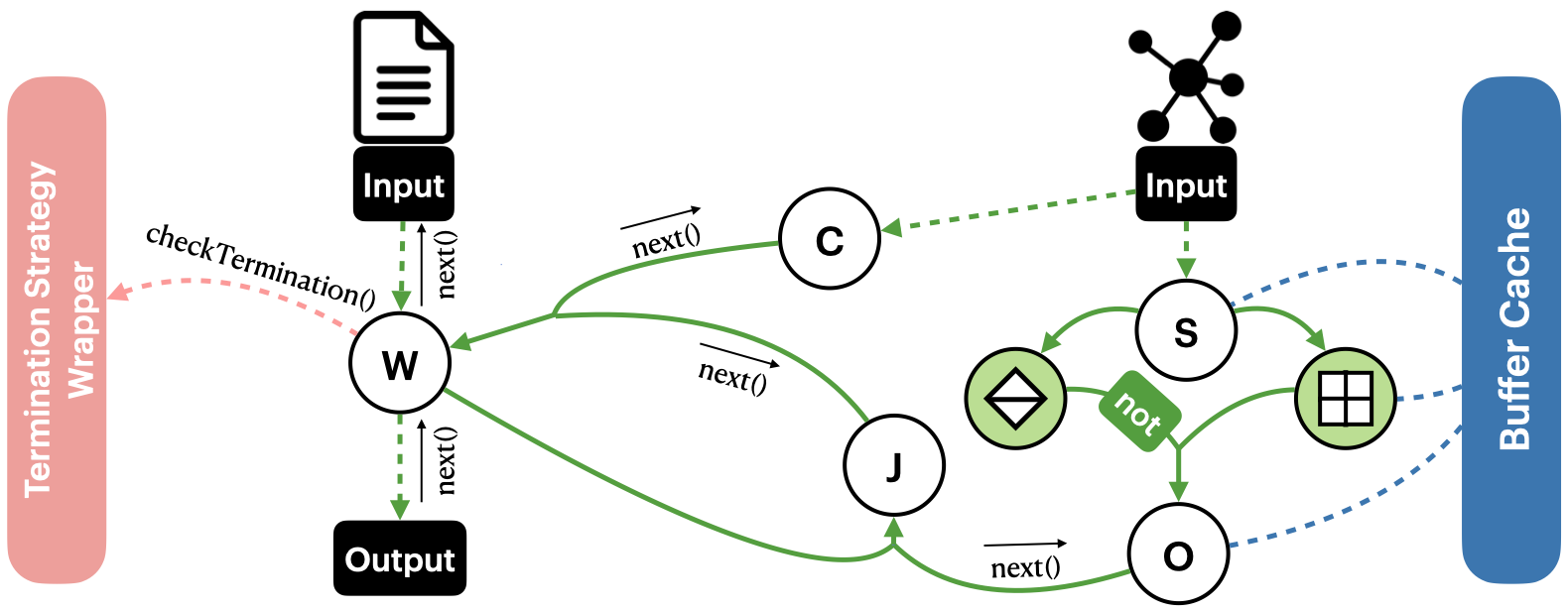}
    \caption{\small{The reasoning pipeline for Example~\ref{ex:running_example}. The atom \textit{significantShare} is denoted by the filter \texttt{S}, \textit{significantOwner} by \texttt{O}, \textit{watchCompany} by \texttt{W}, \textit{connected} by \texttt{C}, and \texttt{J} is an artificial filter to decompose, for simplicity, the ternary join of Rule~2 into binary joins.}}
    \label{fig:execution-schema}
\end{figure}

\subsection{A Time-aware Execution Pipeline}
\label{sec:execution-pipeline}

Similar to the \emph{pipe and filters} architecture~\citep{BuHS07}, a DatalogMTL program $\Pi$ is transformed into an execution pipeline that takes data from the input sources, applies necessary modifications, including relational algebra operations (e.g., projection, selection) or time-based ones, and generates the intended output as a result.

\smallskip\noindent
\textbf{Building the pipeline}
involves four main steps: (i) A \textit{logic optimizer} rewriting tasks to simplify programs and ensure that only combined individual operators are allowed in the canonical form. (ii) A \textit{logic compiler} transforming
the rules into in-memory placeholder objects, each assigned the ``task of knowing'' the transformation to be performed. (iii) A \textit{heuristic optimizer} producing
variations of the generated pipeline to improve performance by introducing perturbations and ad-hoc simplifications. (iv) Finally, a \textit{query compiler} translating the logical graph structure into a \textit{reasoning query plan}. Each placeholder generates a filter responsible for performing the transformations; each read-write dependency between rules induces a pipe. Figure~\ref{fig:execution-schema} shows the Example~\ref{ex:running_example} pipeline.

\smallskip\noindent
\textbf{At runtime,}
a \textit{pull-based} approach is used: sinks iteratively pull data using \texttt{next()} and \texttt{get()}, propagating calls through a filter chain to source filters. Source filters read from the initial data source via data adapters. Each filter applies transformations based on rule types (linear, joins, temporal operators, aggregations, etc.).
\Revision{In particular, a linear filter handles \RevisionTwo{Vadalog linear rules, defined as rules having a single atom in the body\footnote{\RevisionTwo{Different definitions of 'linear rule' exist in the literature, which can lead to ambiguity when comparing works and perspectives. To avoid confusion, we will refer to our version as the
``Vadalog linear rule''.}}.}} As long as data is available in the cascade of invoked filters, 
\texttt{next()} succeeds. 
In Figure~\ref{fig:execution-schema}, for example, facts for the output filter \texttt{W} are generated both with the input data \texttt{C} and \texttt{S} and recursively through the filter \texttt{J} since the \textit{watchCompany} rule is recursive.

\smallskip\noindent
\textbf{Temporal Challenges}. We will now give a full overview of the system, pointing out how Temporal Vadalog provides support for tackling a number of time-relevant challenges.

\begin{itemize}   
    \item \textbf{Applying Temporal Operators.} Section~\ref{sec:temporal-operators} covers how the temporal operators (e.g., the $\diamondminus$ operator in Figure~\ref{fig:execution-schema}) are implemented in the pipeline.  
    \item \textbf{Merging strategies.} The semantics for $\boxminus$ requires the merging of adjacent and overlapping intervals to function correctly. How merging strategies work and are designed in Temporal Vadalog is covered in Section~\ref{sec:architecture_merge_strategies}. 
    \item \textbf{Temporal Joins and Stratified Negation}. Section~\ref{sec:architecture_join} explains the implementation of joins in temporal reasoning (e.g. in the \texttt{W} filter), where time-awareness, as well as support for stratified negation, is required.
    \item \textbf{Termination Strategies}. DatalogMTL
    can formulate programs with
    infinite models~\citep{BeNS21}, by capturing domain events that repeat infinitely, such as weekdays. Section~\ref{sec:termination-infinite} outlines our strategy for program termination.
    \item\textbf{Aggregate functions and numeric operations.} The Temporal Vadalog System provides standard scalar and temporal arithmetic operations, as well as support for aggregate functions in the form of time-point or cross-time monotonic aggregations. This unique feature allows for a non-blocking implementation that works seamlessly with recursion. As far as we know, the Temporal Vadalog System is the only DatalogMTL reasoner capable of implementing aggregations. Their syntax and semantics are explained in our recent publication~\citep{DBLP:conf/ruleml/BellomariniNS21}.

    \item \textbf{Temporal and non-Temporal Reasoning}. 
Since many existential extensions are undecidable~\citep{DBLP:conf/datalog/LanzingerW22}, we consider temporal reasoning and existential reasoning as orthogonal fragments: in Datalog with existentials~\citep{BeBS20} we forbid temporal operators and in DatalogMTL we forbid existentials. To support both modes within one program we added support for temporal wrapping and unwrapping of rules by allowing to switch between intervals as metadata and as atoms.
    
    \item \textbf{Time Series}. The handling of time series and the most common operators in this kind of analysis will be presented in Section~\ref{sec:time-series}.
\end{itemize}
\subsection{Temporal Operators}
\label{sec:temporal-operators}

The implementation of the temporal operators introduced by DatalogMTL is designed in a two-phase procedure, first by applying logical optimization to reduce the number of required rules, and then by introducing multiple filter nodes in the reasoning pipeline.

\paragraph{Logical Optimization.} 
\Revision{The application of linear temporal operators, that is, the operators for \RevisionTwo{Vadalog linear rules}, \emph{box} and \emph{diamond}, results in intervals computed through mathematical expressions between the interval of a fact, $\langle t_1, t_2 \rangle$, and that of the operator itself, $\varrho = \langle o_1, o_2 \rangle$, as schematized by~\cite{WCKK19}:
{\small
\begin{minipage}{.45\linewidth}
  \centering
\begin{align*}
    \langle t_1 + o_1, t_2+o_2 \rangle \tag{$\diamondminus$}\\
    \langle t_1 + o_2, t_2+o_1 \rangle \tag{$\boxminus$} 
\end{align*}
\end{minipage}%
\begin{minipage}{.45\linewidth}
  \centering
\begin{align*}
    \langle t_1 - o_2, t_2-o_1 \rangle \tag{$\diamondplus$} \\
    \langle t_1 - o_1, t_2-o_2 \rangle \tag{$\boxplus$}
\end{align*}
\end{minipage}
\\}
with corresponding rules for determining the interval boundaries.\footnote{\Revision{To be concise, we only see examples with closed intervals, where the application of temporal operators does not change them. 
The transformation tables can be found in~\cite{WCKK19}}.}\\ 
Among these, $\diamondminus$ offers one of the most accessible intuitions for understanding how an interval is derived: 
let us say that a shop has \textit{recentlyOpened} if it had its inauguration in the past 12 days: \RevisionTwo{$\diamondminus_{[0,12]} \mathit{inauguration}(X) \to \mathit{recentlyOpened}(X)$.} Assume that shop \textit{A} has had its inauguration on the 5th and 6th days of the month: \RevisionTwo{$D = \{\mathit{inauguration}(A)@[5,6]\}$.} Then, \textit{A} can be considered to have \textit{recentlyOpened} between the 5th and the 18th: $\mathit{recentlyOpened}(A)@[5,18]$, that is $[t_1 + o_1, t_2+o_2]$. \\ 
As these expressions can be composed, one can rewrite a chain of \RevisionTwo{Vadalog linear rules} containing temporal operators\footnote{\Revision{The logical optimization does not consider $\SinceOperator$ (\emph{Since}) or $\UntilOperator$ (\emph{Until)} as these operators are not linear.}} into a single rule with an equivalent temporal operator, as long as every intermediate step of the chain yields a valid (i.e., non empty) interval. 
} 
\\
\Revision{Since the derived interval can include both positive and negative endpoints, an equivalent operator cannot be chosen among the linear operators $\diamondminus$, $\diamondplus$, $\boxminus$, $\boxplus$, and instead is implemented as a generic operator $T\langle e_1,e_2 \rangle$ where $\langle e_1,e_2 \rangle$ is the desired interval transformation. That is, $T\langle e_1,e_2 \rangle$ applied to the interval $\langle t_1, t_2 \rangle$ of a fact gives $\langle t_1 + e_1, t_2 + e_2 \rangle$. Note that $T$ is not constrained by the usual restrictions for intervals in linear operators, i.e., $e_1$ and $e_2$ are arbitrary for $T$. The linear operators $\diamondminus$, $\diamondplus$, $\boxminus$, $\boxplus$ can easily be seen as instances of the generic $T$ operator: we have that $T\langle e_1,e_2 \rangle$ expresses any of the following: $\diamondminus_{\langle e_1,e_2 \rangle}, \boxminus_{\langle e_2,e_1 \rangle}, \diamondplus_{\langle -e_2,-e_1 \rangle}$, or $\boxplus_{\langle -e_1,-e_2 \rangle}$. }

\smallskip\noindent
\Revision{\textbf{Algorithm.} Given a chain of \RevisionTwo{Vadalog linear rules}, we apply the mathematical expressions of each operator in order, starting from $[0,0]$, the smallest valid interval for facts. If the resulting interval is empty---that is, the left endpoint is larger than the right endpoint, or, in case it is equal, the interval is open---, or if we reach the end of the chain, we stop and merge the rules to an equivalent temporal operator applying the derived interval. This process is applied exhaustively until all rules are merged.} 

\begin{example}
\textit{Consider the following combination of three temporal operators, where (\Revision{Expression}~3) can be reduced to $T[9,9]$, while (\Revision{Expression}~4) can be reduced only partially.}
{\small
\begin{align}
    \boxminus_{[5,10]} \diamondminus_{[2,5]} \diamondplus_{[1,3]} A \tag{3}\\
    \diamondminus_{[2,5]} \boxminus_{[5,10]} \diamondplus_{[1,3]} A \tag{4}
\end{align}}
\Revision{For Expression~3, we first apply $\diamondplus_{[1,3]}$\footnote{\Revision{Note that the operators are applied from inside to outside to the given fact.}} to $[0,0]$, resulting in $[-3,-1]$, then apply $\diamondminus_{[2,5]}$, resulting in the interval $[-1,4]$. Then, we combine this interval with the remaining operator $\boxminus_{[5,10]}$, resulting in $[9,9]$ and in the final expression $T[9,9] A$. For Expression~4, we again apply $\diamondplus_{[1,3]}$ to $[0,0]$, resulting in $[-3,-1]$, but then apply $\boxminus_{[5,10]}$ resulting in the interval $[7,4]$. As no further optimization is possible, the final expression is $\diamondminus_{[2,5]} T{[7,4]} A$.}
\end{example}

\paragraph{Reasoning pipeline.} A filter node is introduced for each operation executed in the pipeline.
For the operators introduced by DatalogMTL, we consider the following:
\begin{itemize}
    \item \textit{TemporalNode.} \Revision{This node computes the resulting interval of a fact at the application of a linear temporal operator ($\diamondminus$, $\diamondplus$, $\boxminus$, $\boxplus$, or the generic $T$ operator introduced in \emph{Logical Optimization}, following the DatalogMTL semantics as shown above).}
    \item \textit{MergeNode.} 
    This node merges overlapping and adjacent intervals, required for the application of the box operator. Its functioning is addressed in Section~\ref{sec:architecture_merge_strategies}. 
    \item \textit{ClosingNode.} This node \Revision{computes the closure of} the interval of a fact. It is required before the application of $\SinceOperator$ and $\UntilOperator$, \Revision{as defined in~\citep{WCKK19}: $\varrho_{\SinceOperator} = ((\varrho_1^c \cap \varrho_2) + \varrho_3) \cap \varrho_1^c$, resp. $\varrho_{\UntilOperator} = ((\varrho_1^c \cap \varrho_2) - \varrho_3) \cap \varrho_1^c$ where $c$ denotes the closing operator and $+$ (resp. $-$) is the addition (subtraction) of intervals as defined for $\diamondplus$ ($\diamondminus$). } 
    \item \textit{TemporalJoinNode.} An extension of the join node in Vadalog incorporating restrictions on intervals of facts, required for $\SinceOperator$ and $\UntilOperator$ to take the semantics of the special join behavior of intervals into account. We discuss this node in detail in Section~\ref{sec:architecture_join}.
\end{itemize}

\subsection{Merging Strategies}
\label{sec:architecture_merge_strategies}
This section discusses the implementation of the MergeNode, which provides us with two orthogonal optimization choices: 
(i) the placement strategy and (ii) the merge strategy. 
\begin{example}
\label{ex:merge_1}
\textit{A person $X$ that owns an amount $Z$ of shares of a company $Y$ is an investor of \Revision{the} company (Rule~5), and a long-time investor of \Revision{the} company if it is an investor for at least 3 years with gaps of at most half a year (Rule~6).
{\small
    \begin{align*}
        \mathit{shares}(X,Y,Z) & \to \mathit{investor}(X,Y) \tag{5}\\
        \boxminus_{[0,3]}  \diamondminus_{[0,0.5]} \mathit{investor}(X,Y) & \to \mathit{longTimeInvestor}(X,Y) \tag{6}
    \end{align*}}
    with the following database:
{\small
    \begin{align*}
        \mathcal{D} = \{\mathit{shares}(A,B,0.2)@[0.1,0.5), \mathit{shares}(A,B,0.2)@[0.4,1.1), \\ 
        \mathit{shares}(A,B,0.3)@[1.5,3.7), \mathit{shares}(A,B,0.4)@[3.7,4.2)\}
    \end{align*}
    }
}
\end{example}

\paragraph{Placement Strategy.}
The system maintains an ordered data structure of intervals for each fact.
Depending on the program, there are \Revision{multiple} ways to merge facts to uphold correctness. Consider Example~\ref{ex:merge_1}, where one can merge directly before the box operator (minimal merge), before the diamond operator (earliest merge), or continuously, i.e., always guaranteeing that all intervals per fact are merged (always merge) throughout the reasoning process. 
The placement options are described as follows:
\begin{itemize}
    \item \textit{Minimal Merge.} The planner inserts a merge operation before every box operator. This ensures correctness but does not provide further performance optimization. 
    \item \textit{Earliest Merge.} The planner inserts the merge in the earliest position it can be processed in the pipeline, so that the intervals of each fact are already merged when it reaches the box operator, while the operators coming before the latter also benefit from the reduced number of facts. If no merge operation is required (i.e., there is no box operator) the planner avoids merging.
    \item \textit{Always Merge.}   \RevisionTwo{Merge operations are placed after the input filters and all filters that can generate facts with unmerged intervals.}
    This can reduce the number of intermediary facts, providing \RevisionTwo{a memory usage improvement}, yet has a tradeoff with the time required for merging all existing facts.
\end{itemize}
In addition, for the \textit{Minimal} and \textit{Earliest Merge}, further optimization is obtained by providing the planner hints on where to place additional merge nodes, such as: \Revision{after rules that share the same head;} after a diamond operator produces many overlapping intervals; after the input or before the output to eliminate duplicates.

\begin{figure}[t]
    \centering
\includegraphics[width=0.67\textwidth]{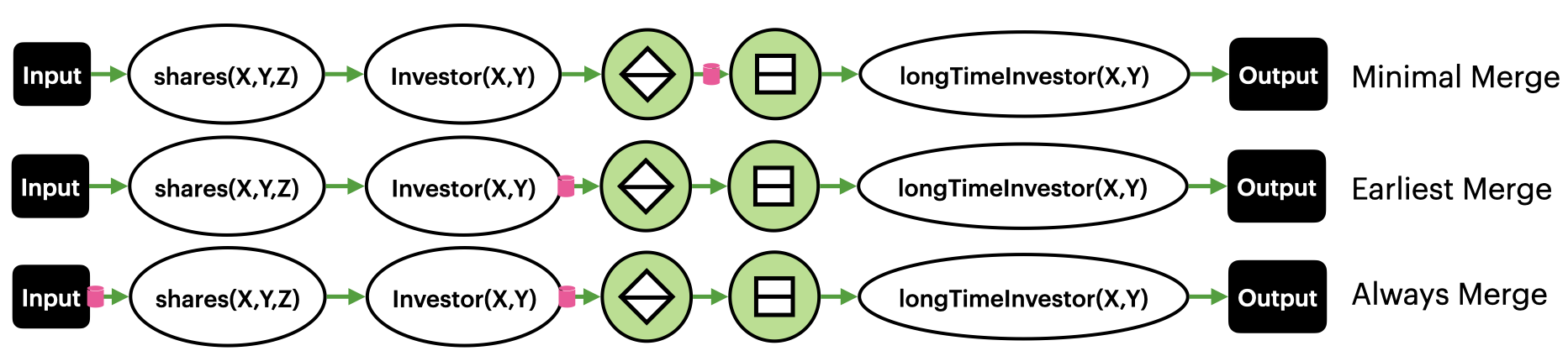}
    \caption{\small{Overview of interleaving strategies; merging positions are marked in fuchsia.}}
    \label{fig:interleaving_strategies}
\end{figure}

Figure~\ref{fig:interleaving_strategies} shows the merging strategies for Example~\ref{ex:merge_1}. 
%Note that using temporal operators will always transform any set of intervals into a merged set, as coalescing is always applied before the output. 
Temporal operators transform any set of intervals into a merged set: coalescing is always applied before the output.

\paragraph{Merge Strategy.}
\begin{algorithm}[t]
\small
\caption{\RevisionEnvTwo{Streaming Strategy in the MergeNode}}
\label{alg:merge_streaming_strategy}
\begin{algorithmic}[1]
\State $ \mathrm{mergeStructure} \gets \mathit{createMergeStructure}()$
\Function{Next}{}
\State $ \mathrm{next} \gets \RevisionEnvTwo{\mathit{Linear.next}()}$
\If{$ \mathrm{next} $}
\State $(\mathrm{changed}, \mathrm{mergedEntry}) \gets \mathrm{mergeStructure}.\mathit{add}(\RevisionEnvTwo{\mathit{Linear.get}()})$
\State $\RevisionEnvTwo{\mathit{set}(\mathrm{mergedEntry})}$
\EndIf
\State \Return $ \mathrm{next} $
\EndFunction
\end{algorithmic}
\end{algorithm}
\begin{algorithm}[t]
\small
\caption{\RevisionEnvTwo{Blocking Strategy in the MergeNode}}
\label{alg:merge_blocking_strategy}
\RevisionEnvTwo{
\begin{algorithmic}[1]
\State $ \mathrm{mergeStructure} \gets \mathit{createMergeStructure()}$
\State $ \mathrm{currentPosition} \gets 0$
\Function{Next}{}
\State $\mathrm{changed} \gets \mathit{false}$
\If{$ \mathrm{currentPosition} \geq \mathrm{mergeStructure}.\mathit{length}$}
\While{$\RevisionEnvTwo{\mathit{Linear.next}()}$}
    \State $\RevisionEnvTwo{(\mathrm{newChange}, \mathrm{mergedEntry}) \gets \mathrm{mergeStructure}.\mathit{add}(\mathit{Linear.get}())}$
    \State $\RevisionEnvTwo{\mathrm{changed} \gets \textrm{changed or newChange}}$
\EndWhile
\EndIf
\If{$ \mathrm{changed} $}
    \State $  \mathrm{currentPosition} \gets 0$
\EndIf
\State \RevisionEnvTwo{$\mathit{set}(\mathrm{mergeStructure}.\mathit{getAtPosition}(\mathrm{currentPosition}))$ \Comment{\RevisionEnvThree{set() automatically handles nulls}}}
\State $ \RevisionEnvTwo{ \mathrm{currentPosition} \gets  \mathrm{currentPosition} + 1}$
\State \Return $  \mathrm{currentPosition} < \mathrm{mergeStructure}.\mathit{length}$
\EndFunction
\end{algorithmic}}
\end{algorithm}
In software development, we face a typical trade-off between streaming and blocking-based processing operations: the former typically aims for system responsiveness, limited memory footprint, and in-memory computation, while the latter aims for improving overall performance but requires more memory or materialization of intermediate results~\citep{Scio20}.
In the case of the merging operator, we recognize a similar behavior. 
The merging operator is partially blocking: it can return a partial merge, but has to wait for all facts to be processed to return the final merged intervals. 
Two implementation options, \Revision{\textit{Streaming} and \textit{Blocking},} were
integrated into the system.
\Revision{Algorithms~\ref{alg:merge_streaming_strategy}-\ref{alg:merge_blocking_strategy} present
%the logic for 
these strategies through the function \textsc{next()}, which returns a Boolean denoting whether a new fact has been retrieved (the caller will later retrieve the fact itself through a separate getter function \RevisionTwo{\textsc{get()}}). It makes use of \RevisionTwo{\textit{Linear.next()}}, inherited from the linear filter scan, to check for new facts in the pipeline; thus it retrieves the facts, merges their intervals, and saves them into a \textit{mergeStructure}, a hash map with the \RevisionTwo{ground atoms} as keys and the collections of intervals as values.\footnote{Intervals are stored in a tree-like structure
that merges those adjacent and overlapping automatically.} \RevisionThree{Every node (and hence, every MergeNode) handles only one atom.} 
Updating \textit{mergeStructure} via
\textit{add} also returns a Boolean \RevisionThree{\textit{newChange}} to note whether \textit{mergeStructure} has been updated (i.e., it added new facts or \RevisionTwo{new intervals not subsumed by already-present intervals}.)}

\Revision{\noindent\textit{Streaming.} (Algorithm~\ref{alg:merge_streaming_strategy}) This operator pulls a fact from the previous node, merges the fact on-the-fly with the intermediate merging result, and forwards the merged fact without waiting until all incoming data has been processed.}

\Revision{\noindent\textit{Blocking.} (Algorithm~\ref{alg:merge_blocking_strategy}) This operator receives and merges all available facts before forwarding them\RevisionTwo{: \textit{currentPosition} keeps track of the facts already served downstream.} Newly received facts\footnote{We have to be aware of changes in case the node is part of a cyclic structure.} are only retrieved if the facts in \textit{mergeStructure} have already been handled \RevisionTwo{(Line~5)}; \RevisionTwo{\textit{currentPosition}} is reset \RevisionThree{to 0} if \textit{mergeStructure} is \textit{changed}: \RevisionThree{as we do not know which facts have been changed\footnote{\RevisionThree{This behavior has been chosen as a trade off between space and frequency of occurrence.}}, all of them will have to be served again downstream} (Lines~9-10). \RevisionThree{With \textit{set()} (Line~11) we specify which fact will be retrieved with \textsc{get()} by the \textsc{NEXT()} call from the filter below (an incremental value as long as the \textit{next()} is called and no new facts are retrieved, 0 in case we have to start again because the structure changed).} \RevisionTwo{The calls are performed as long as there are new facts to pull from the filter above. By construction, as shown in Section~\ref{sec:execution-pipeline}, our termination strategies ensure termination also in the presence of cyclicity. Note that the possible merges of the facts have already been performed during the execution of the pipeline and no outstanding merges are left after termination.}}
\begin{example}
\label{ex:merging_example_variation}
\textit{Example~\ref{ex:merge_1} (continued) with the merge operator placed directly before the box. The \emph{Streaming Merge} reads the
first three entries, sufficient to apply $\boxminus_{[0,3]}$ 
to derive the intermediate result \emph{longTime\-Investor(A,B)@\-[3.1,4.2]}, before the final \emph{longTime\-Investor(A,B)@\-[3.1,4.7]} is derived; the \emph{Blocking Merge} strategy 
waits for all facts to be read
then applies the box operator,
deriving only fact \emph{longTime\-Investor(A,B)@\-[3.1,4.7]}.}
\end{example}

\subsection{Temporal Joins}
\label{sec:architecture_join}

\Revision{\begin{example}
\label{ex:temporal_join}
\textit{A person $X$ that goes \RevisionTwo{to} a movie matinee $M$ (between $[14,16)$) gets a discounted ticket \RevisionTwo{(Rule~7 and database)}. In the example, only person $A$ will get a discount.
{\small
\begin{align*}
        \mathit{goesToTheMovies}(X,M), \mathit{matineeDiscount}(M) \to \mathit{discountedTicket}(X) \tag{7}\\
\mathcal{D} = \{\mathit{goesToTheMovies}(A,C)@[15,17), \mathit{goesToTheMovies}(B,C)@[21,23), \\ 
        \mathit{matineeDiscount}(C)@[14,16)\}
\end{align*}
}
}
\end{example}}
\begin{algorithm}[t]
\small
\caption{Temporal Join between two predicates}
\textbf{Input:} predicates $A_0$ and $A_1$  to be joined, \RevisionEnvTwo{with $A_0$ not negated}
  \RevisionEnv{
\begin{algorithmic}[1]
    \State $I_0 \gets A_0.iterator()$
    \State $I_1 \gets A_1.iterator()$
    
    \State $(a_0, \varrho_0) \gets I_0.getNext()$
    \Function{Next}{}
        \State $interval \gets \varrho_0$
        
        \While{true}
            \State \RevisionTwo{$ X \gets a_0.joinTerm$}
            \State \RevisionTwo{$(a_1, \varrho_1) \gets A{_1}.index.get(\mathit{X})$} %\Comment{\LiviaComments{does this mean: look in the precalculated index?}}
            \If{$a_1$ is null} \Comment{Continue full scan, if index miss}
                \While{$I_1.next()$}
                    \State $(a_1, \varrho_1) \gets I_1.getNext()$
                    \State $A{_1}.index.put(a_1)$ \Comment{Update the index map for $A_1$}
                    \If{$a_1.joinTerm == a_0.joinTerm$} %\Comment{\LiviaComments{how to better explain here?}}
                        \State \textbf{break} \RevisionTwo{\Comment{Exit the inner loop if matching fact is found}}
                    \EndIf
                \EndWhile
            \EndIf
            \If{$a_1$ is null} \Comment{No further matching fact $A_1$ found}
                \If {$A_1$ is negated}
                    \State \textbf{return} true
                \EndIf
                \If {$I_0.next()$ is false}
                    \State \textbf{return} false
                \EndIf 
                \State \RevisionTwo{$(a_0, \varrho_0) \gets I_0.getNext()$}  \Comment{Repeat loop for next $a_0$}
                \State $interval \gets \varrho_0$
                \State \textbf{continue}
            \EndIf
            \State $interval \gets resultingIntervalFromJoinLogic(\varrho_1, interval)$
            \If {$interval$ is empty} 
            \If {\RevisionEnvThree{$A_1$ is negated}}
                \If {$I_0.next()$ is false}
                    \State \textbf{return} false
                \EndIf 
                \State $(a_0, \varrho_0) \gets I_0.getNext()$  \Comment{Repeat loop for next $a_0$}
                \State $interval \gets \varrho_0$
            \EndIf
            \State \textbf{continue}
            \EndIf
            \If {$A_1$ is not negated}
                \State \textbf{return} true
            \EndIf
        \EndWhile
    \EndFunction
\end{algorithmic}}
\label{alg:temporal_joinsrev}
\end{algorithm}

\noindent
\Revision{When joining two or more temporal predicates, we are interested in joining them not only based on their terms, but also on the temporal interval of each fact. We call this particular join a ``temporal join''.}

The \Revision{temporal} join in Temporal Vadalog is an extension with intervals of the Vadalog slot machine join~\citep{BeBS20}, an enhanced version of the index nested loop join~\citep{DBLP:books/daglib/0020812} with the support of dynamic in-memory indexing: instead of having a pre-calculated index, it builds the index in-memory during the first full scan. 
Algorithm~\ref{alg:temporal_joinsrev} shows the operation for joining two predicates ($n=2$). 
For each predicate $A_{k}$ with $0 \leq k < n$ to be joined, we first use the index to get the next scanned fact that matches the known terms \Revision{used in the join, i.e., the \textit{joinTerm}} from the previous $A_j$ with $0 \leq j  < k$ (Line 8). If no further fact is found in the index, we do the full scan (Lines \RevisionTwo{10-14}) until either a matching fact is found or the number of facts is exhausted.
%In case 
If no further fact is found (Line \RevisionTwo{15-22}), we continue the scan with the next $A_{j}$ if it exists\RevisionTwo{ and the current $A_k$ is not negated.}
In case a fact has been found\RevisionTwo{ (whether from the index or the full scan)}, we update the valid interval of the joined fact depending on the join logic (Line \RevisionTwo{23}): 
the difference for a negated literal, the intersection for a positive interval, or a mixture of interval operations and set operations for $\UntilOperator$ and $\SinceOperator$.
Finally, we check whether the resulting interval is valid and if not, \RevisionThree{if $A_{k}$ is negated, we continue the scan with the next $A_{j}$ if it exists (Line \RevisionThree{25-29}), otherwise we continue with the loop}; in case \RevisionThree{the interval is not empty and $A_{k}$ is not negated}, we return true as we have found a valid joined fact (Line \RevisionThree{31-32}); otherwise, we continue to retrieve the next ``negated'' fact.

\RevisionTwo{As the temporal join supports negated atoms, it follows that Temporal Vadalog supports stratified negation \RevisionThree{in safe rules}. }

\subsection{Termination Strategy for the Infinite Chase of Intervals}
\label{sec:termination-infinite}

In previous work~\citep{BeNS21}, we discussed the fragments of DatalogMTL that can generate infinite models that have finite representations in DatalogMTL$^{\mathit{FP}}$. In the following example, we depict one such case. This section explains how Temporal Vadalog handles these cases to guarantee termination.

\begin{example}
\textit{A \Revision{30-day} Job Report is an economic indicator, released once every \Revision{30 days}, whose content can impact stock prices. 
Stock market analysts want to be aware of this indicator's releases to mark them as possible causes of stock price change
\Revision{over a certain threshold ($K$ below)}.
\label{ex:termination-example}
{\small
    \begin{align*}
        \diamondminus_{[30,30]}\mathit{30DayJobReport} &\to \mathit{30DayJobReport} \tag{8}\\
        \mathit{StockPriceChange}(X,V), V>K &\to \mathit{PriceEvent}(X) \tag{9}\\
        \mathit{PriceEvent}(X), \diamondminus_{[0,1]} \mathit{30DayJobReport} &\to \mathit{PossibleCause}(X, \mathrm{``JR"})\tag{10}\\ 
        D=\{\mathit{30DayJobReport@[0,0]},    \mathit{StockPriceChange}&(A,K+1)@[121,121], ... \}
    \end{align*}}
}    
\end{example}

\noindent
A DatalogMTL$^{\mathit{FP}}$ (resp. BP) program can fall into one of three categories~\citep{BeNS21}: (i) it is \textit{harmless}, i.e., it satisfies the sufficient conditions for a finite model; (ii) it is either DatalogMTL$^{\mathit{FP}}_{\diamondminus}$ \textit{temporal linear} or DatalogMTL$^{\mathit{FP}}_{\boxminus}$ \textit{union-free}, i.e., it satisfies the sufficient condition for a constant model under certain constraints; (iii)  
%if the program does not fall under the previous two categories, 
otherwise, it belongs to the DatalogMTL$^{\mathit{FP}}$ category, a sufficient condition for \RevisionTwo{a \textit{periodic}} model.

The Temporal Vadalog system guarantees termination by employing a two-phase approach, one at \textit{compile time} and one at \textit{runtime}.

\smallskip\noindent\textbf{Compile time.} In this phase, the planner detects the fragment of DatalogMTL$^{\mathit{FP}}$: 
\begin{enumerate}[label=\alph*)]
    \item The planner checks if the program has ``harmful'' temporal cycles using~\cite[Algorithm~1]{BeNS21}. If the program is \textit{harmless}, it marks $\textit{modelKind}=\textit{Finite}$.
    \item In presence of
    harmful temporal cycles, the algorithm checks if the program is \textit{temporal linear}, that is, each rule has at most one body predicate mutually temporal recursive with the head in the dependency graph of $\Pi$. If the program is temporal linear and \Revision{the operators allowed in temporal linear, defined over} $[t1,t2]$ are such that $t1 \neq t2$, then the algorithm sets the $\textit{modelKind}=\textit{Constant}$.
    \item If the program is not temporal linear, the algorithm checks if the program is \textit{union-free}. A program is union-free if there are no rules of $\Pi$ sharing the same head predicate. If the program is union-free and the box operators $\boxminus_{[t1,t2]}$ are such that $t1 \neq t2$, then the algorithm sets the $\textit{modelKind}=\textit{Constant}$.
    \item If the program does not meet the conditions to be temporal linear or union-free, we assume it is in DatalogMTL$^{\mathit{FP}}$ and set  $\textit{modelKind}=\textit{Periodic}$.
\end{enumerate}

\noindent
At this time, the system computes the repeating pattern length \textit{pLength} in all non-finite cases, based on the pattern lengths combination of the different Strongly Connected Components (SCC) of $\Pi$. The resulting facts are of the form  $P(\tau)@\varrho$ and $\{P(\tau)@\langle o_1, o_2 \rangle, \mathit{n}\}$, while the intervals are given by $\langle o_1 + \mathit{x} * \mathit{pLength} , o_2 + \mathit{x} * \mathit{pLength} \rangle$ for all $x \in \mathbb{N}$, where $x \geq \mathit{n}$, in the case of \RevisionTwo{\textit{periodic}}, or $\langle o_1 + \mathit{x} * \mathit{pLength} , \infty \rangle$, in the case of \RevisionTwo{\textit{constant}}.
Functional components called \textit{termination strategies} wrap all filters in the pipeline, \Revision{to prevent the generation at runtime of facts leading to non-termination.} 

%the generation of specific facts at runtime that could lead to non-termination. 

\smallskip\noindent\textbf{Runtime.} 
In this phase, the system behavior depends on the detected fragment (\Revision{the one associated with the set} \textit{modelKind}), 
denoting \textit{fragment awareness}. 
If \textit{modelKind} is \textit{finite}, 
non-termination can only be caused by Datalog recursion, and standard termination strategies are employed.
The reasoning produces facts of the form $P(\tau)@\varrho$. If the model is \textit{constant} or \textit{periodic}, the termination strategies recognize facts generated by the ``non-finite'' filters and mark 
the matching repeating patterns.
In the case of \RevisionTwo{a \textit{constant}} model, the base interval of ground atoms is converted into  $\langle o_1 + \mathit{x} * \mathit{pLength}, \infty \rangle$, and the generation stops. If the model is \RevisionTwo{\textit{periodic}}, the termination strategies related to the ``non-finite'' filters represent the numeric intervals with the equivalent of their pattern-based symbol, so that repeating sub-intervals are not generated, and termination is guaranteed. 

\smallskip Looking back at Example~\ref{ex:termination-example}, at compile time the planner  determines that $\textit{modelKind}=\textit{Periodic}$, and, from Rule~10, that $pLength=1$. At runtime, after fact \Revision{$\mathit{30DayJobReport}@[60,60]$} is generated, the termination strategy for \textit{PossibleCause} infers that $n=0$ and so that all facts generated from Rule~8 have form $\mathit{30DayJobReport}@[x\times 30,x\times 30 + 1]$ for $x \ge 0$, and their generation is put on hold. Applying the join for $\mathit{PriceEvent}@[121,121]$ and the pattern generated from Rule~8-Rule~10, $x \in \mathbb{N}$ such that $[x\times 30,x\times 30 + 1] \cap [121,121]$ is not an empty interval and corresponds to $x=4$. We can conclude that $\textit{PossibleCause}(A,\mathrm{``JR"})@[121,121]$.

\section{Time Series in Temporal Vadalog}
\label{sec:time-series}

Time series analysis is vital across many sectors, using historical data to identify trends, detect anomalies, predict events, and inform decisions to improve efficiency, reduce costs, and allocate resources.
While many Time Series Database (TSDB) systems exist, 
considering time series analysis relevance in data science,
we want to explore how to handle it 
with a more general-purpose system like Temporal Vadalog.
allowing us to use common operators for time series in the broader scope of reasoning applied, for instance, to KGs, specifically exploiting its characteristics of explainability and context-awareness, especially important 
for domains dealing with sensitive data and particular needs. 

In this section, we will present how the Temporal Vadalog System can be used to reason over time series, by showing how many of the core functions of time series databases are easily and immediately available by using DatalogMTL operators, monotonic temporal aggregations and arithmetic operations. 
Since we focus on regular time series, the examples in the following sections are intended to be used in discrete time (e.g., timestamps).

\subsection{Basic operations}
Temporal Vadalog natively handles the \textit{sum} and other arithmetical operations over the same time intervals. This section covers other basic operations that deal with time series: shifting, rolling, and resampling. We will proceed informally by example, as the temporal operators' semantics has already been introduced.

\begin{figure}[t]
    \centering
    \includegraphics[width=0.87\textwidth]{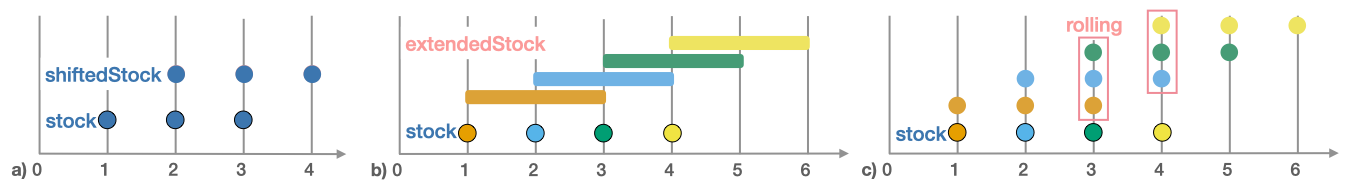}
    \caption{\small{Basic time series operators in Temporal Vadalog: 
    a) Shifting by 1; 
    b) Rolling operator with $n=3$;
    c) Join of the extended intervals with the original ones}}
    \label{fig:shifting-rolling}
\end{figure}
\medskip\noindent \textbf{Shifting.} 
Shifting re-positions the time points of a time series by adding a delay, 
achieved through the box or diamond operator by stating a fact holds at $t$ if it held at some past time point.
The example adds a lag of one unit of time to the stock facts, where $\mathit{stock}(X,\mathit{Value})$ represents any time series 
$X$ with value $\mathit{Value}$.
Figure~\ref{fig:shifting-rolling}a shows the shifting operation over a discrete timeline.
\begin{align*}
  \diamondminus_{[1,1]} \mathit{stock}(X, \mathit{Value}) & \to \mathit{shiftedStock}(X, \mathit{Value}) \tag{11}
\end{align*}
 
\medskip\noindent \textbf{Rolling.} 
Rolling defines a fixed-size window that slides through a time series one data point at a time, allowing for the computation of statistics over the data points within the window.
A rolling window of size $n$ is achieved by extending each data point's interval by $[0,n)$ with the diamond operator
(Figure~\ref{fig:shifting-rolling}b). The time series is later joined with the original one, 
applying the full window's data points to each specific interval for further operations (Figure~\ref{fig:shifting-rolling}c where at $t=3,4$ we have the $n=3$ rolling windows.)
\begin{align*}
\diamondminus_{[0,n)} \mathit{stock}(X,\mathit{Value}) & \to \mathit{extended}(X, \mathit{Value}) \tag{12}\\
\mathit{stock}(X, \mathit{Value}), \mathit{extended}(X,Roll) & \to \mathit{rolling}(X,\mathit{Roll}) \tag{13}
\end{align*}

\medskip\noindent \textbf{Resampling.} 
The resampling of a time series changes its time resolution.
%or frequency. 
Working with different time resolutions (e.g., one monthly series and one daily) may require this change in both directions: \textit{downsampling}, lowering the frequency of the data (e.g. daily to monthly), and \textit{upsampling}, increasing the frequency 
(e.g. monthly to daily).  

\smallskip\noindent
\textit{Downsampling.} In Temporal Vadalog we can achieve  \textit{downsampling} by using the aggregation operator $\triangle_{\mathit{unit}}$, where \textit{unit} is the frequency we want to transform our time series into. This operator was introduced in a previous work~\citep{DBLP:conf/ruleml/BellomariniNS21}. 
\begin{align*}
\triangle_{\mathit{month}} \mathit{dailyStock}(X,\mathit{Value}) &\to \mathit{monthlyStocks}(X,\mathit{Value}) \tag{14}
\end{align*}
The set of facts that were valid daily will be now valid over the entire month.
To obtain the final value for the monthly data point, we can use arbitrary aggregations, depending on the domain (e.g., arithmetic mean, minimum, maximum, etc.)

\smallskip\noindent
\textit{Upsampling.} We can convert a time series to a higher frequency by using the temporal join with a series with a higher frequency that we can create on the spot with a diamond operator. In the following example, we will convert a \textit{weeklyStock}\Revision{(Company,Value,WeekStart), where $\mathit{WeekStart}$ is expressed in a day \Remove{time}unit,} into a daily time series.
\begin{align*}
\diamondminus_{[1,1]} \mathit{dailySeries}(X), Y=X+1 &\to \mathit{dailySeries}(Y) \tag{15}\\    
\diamondplus_{[0,7)} \mathit{weeklyStock}(C,V,W) & \to \mathit{nextWeek}(C,V,W) \tag{16}\\
\mathit{dailySeries}(Z), \mathit{weeklyStock}(C,V_1,W_1),  \\ \mathit{nextWeek}(C,V_2,W_2),  
W=W_1 + \tfrac{(V-V_1)(W_2-W_1)}{(V_2-V_1)} & \to \mathit{upsampling}(C,W) \tag{17}
\end{align*}
Note that \RevisionTwo{Rule~15} generates an infinite \Revision{\textsc{chase}} of intervals, which requires us to employ a mechanism for terminating the generations like the strategies discussed in Section~\ref{sec:termination-infinite}.

\subsection{Moving Averages}
\label{sub:mva}
Moving averages use the rolling window described above to calculate different types of averages, typically to smooth out fluctuations and noise.

\begin{figure}[t]
    \centering
    \includegraphics[width=0.63\textwidth]{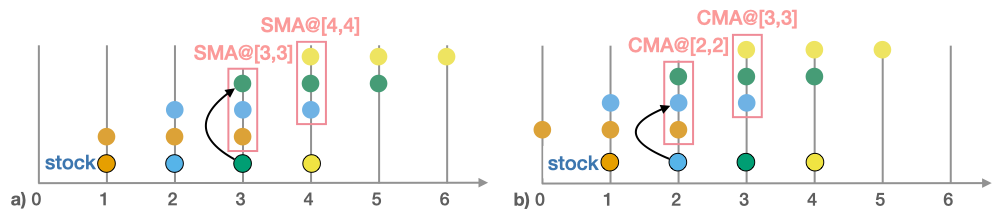}
    \caption{\small{From Simple moving average (a) to Centered moving average (b), with $n=3$.}}
    \label{fig:smacma}
\end{figure}

\medskip\noindent \textbf{Simple Moving Average (SMA).} 
In this moving average, each data point contained in the window has the same weight in the calculation of the average. Hence, we compute first the rolling window and then derive the arithmetic mean of the data points (Figure~\ref{fig:smacma}a).
\begin{align*}
\mathit{rolling}(X,Y), \mathit{Avg}=avg(Y) & \to \mathit{sma}(X,\mathit{Avg}) \tag{18}
\end{align*}

\medskip\noindent \textbf{Exponential Moving Average (EMA).} With the EMA operator we give more importance to the more recent data points.
The first value is calculated through the simple moving average. \Revision{We omit this from the example for conciseness.}  The following values are calculated as a sum (Rule~21) of the current data point multiplied by the parameter $k$ (where $k=2/(1+n)$, $n$ being the size of the window) (Rule~20) and, \Revision{using recursion,} the previous moving average multiplied by $1-k$ (Rule~19). 
\begin{align*}
    \diamondminus_{[1,1]} \mathit{ema}(X), Y = X*(1-k) & \to \mathit{emaInput}(Y) \tag{19}\\
\mathit{stock}(X, V), Y= X * k & \to \mathit{emaInput}(Y) \tag{20}\\
\mathit{emaInput}(X), Y = sum(X) & \to \mathit{ema}(Y) \tag{21}
\end{align*}

\medskip\noindent \textbf{Centered Moving Average (CMA).} In the SMA operator, a window of $n$ data points will yield a value at the right extremity of that window; in the Centered Moving Average (CMA), instead, the data point of reference for the moving average is the data point at the center of the window. 
For any odd $n$ window, this is obtained by calculating the \textit{SMA} and then shifting it by $n/2$, as shown by Figure~\ref{fig:smacma}a-b.
\begin{align*}
    \diamondplus_{[0,n/2)} \mathit{sma}(X,Y) \to \mathit{cma}(X,Y) \tag{22}
\end{align*}

\subsection{Stock and Flow}

Performing mutual conversion between stock and flow is a helpful feature in multiple contexts. Transformation operators, in continuous time, can be computed through derivatives and integrals, while in discrete time one usually considers the difference between two time points. 
We focus on the discrete version, as the discrete (integer) timelines are the usual implementation choice in time series databases~\citep{DBLP:reference/db/Dyreson18}.  

\medskip\noindent \textbf{Stock to Flow (discrete).} This operator calculates the flow as the difference between the stock at $t$ and $t-1$. To do so in Temporal Vadalog, we shift the time series by $[1,1]$ and compute the difference between the values from the original and shifted time series.
\begin{align*}
\diamondminus_{[1,1]} \mathit{stock}(X, V) & \to \mathit{shiftedStock}(X, V) \tag{23} \\
\mathit{stock}(X, V1), \mathit{shiftedStock}(X, V2), V = V1-V2 & \to \mathit{stockToFlow}(X, V) \tag{24}
\end{align*}

\medskip\noindent \textbf{Flow to Stock (discrete).}
This operator, on the other hand, calculates the stock as the cumulative sum of flows from the beginning of the time series up to the current data point. To do so in Temporal Vadalog, we extend the flow time series to the interval $[0,T]$, where $T$ is the last interval, and then we compute the sum over each data point.
{
\begin{align*}
\diamondminus_{[0,T)} \mathit{flow}(X, V) & \to \mathit{rolledFlow}(X, V) \tag{25} \\
\mathit{rolledFlow}(X, V1),
V = sum(V1) & \to \mathit{stock}(X, V) \tag{26}
\end{align*}}

\subsection{Seasonal Decomposition}

Seasonal decomposition separates a time series into three different components: trend, seasonality, and residual. Regular, repeating events such as holidays and seasonal shifts impact variables similarly from one period (e.g. a year) to the next. For this reason, extracting the seasonality from the time series allows having a better understanding of the data.
In this section, we compute seasonal decomposition with the additive model.

\smallskip\noindent
\textbf{Trend}.
The trend component is calculated as the Centered Moving Average (CMA) of the time series, as explained in Section~\ref{sub:mva}, abbreviated in the following as a call to $\mathit{cma}$ with $T$ the length of the period (e.g. 12 months).
\begin{align*}
\mathit{stock}(X, \mathit{Stock}), \mathit{Value} = \mathit{cma}(\mathit{Stock}, \mathit{T}) \to \mathit{trend}(X, \mathit{Value})  \tag{27} 
\end{align*}

\smallskip\noindent
\textbf{Seasonal Component.}
\Revision{The seasonal component is extracted by calculating the per-period averages over the difference between the time series and the \textit{trend} component, which we call \textit{detrend}. In the following example, we calculate the seasonal component over the period $N$ (e.g., 10 years), with $T$ representing the length of the period.}
\begin{align*}
 \diamondminus_{[T, T]} \mathit{detrend}(X, D), \mathit{Sum} = sum(D),  \mathit{Value} = Sum/\mathit{N} & \to \mathit{seasonal}(X, \mathit{Value})  \tag{28}
\end{align*}

\smallskip\noindent
\textbf{Residual.}
Finally, the residual is the time series when the seasonal component and the trend are taken out.
\begin{align*}
\mathit{detrend}(X, D), \mathit{seasonal}(X, S), 
\mathit{Value} = D - S \to \mathit{residual}(X, \mathit{Value})  \tag{29} 
\end{align*}
This concludes our section on time series operations in Temporal Vadalog.

\section{Experiments}
\label{sec:experiments}

For the experimental evaluation of our system, we conducted several performance tests. 
Previously~\citep{BBNS22}, we presented the Company Ownership experiments, a set of five scenarios about ownership changes in companies, whose historical information was stored in Knowledge Graphs of increasing sizes. In this paper, we present new results of the same scenarios, with the optimized reasoning capability of the current Temporal Vadalog and more dynamic datasets (Section~\ref{sub:company-ownership-experiments}), as well as new results from the experiments created with iTemporal~\citep{BellomariniNS22}, an extensible generator of temporal benchmarks, in Section~\ref{sub:itemporal-experiments}. 
We also show the results of new experiments: We replicated the experiments from the MeTeoR paper~\citep{DBLP:conf/aaai/WangHWG22}, the Lehigh University Benchmark ones in Section~\ref{sub:lubm-experiments} and the Meteorological ones in Section~\ref{sub:meteo-experiments}. We also evaluate the performance of reasoning over time series in Section~\ref{sub:timeseries-experiments}.

\smallskip\noindent
\textbf{Setup.} We run the experiments in a memory-optimized virtual machine with 16 cores and 256 GB RAM on an Intel Xeon architecture. Each experiment was run 3 times, and the results shown are the arithmetic mean of the elapsed time of each set of runs. The time out is set at 4,000 seconds ($\sim$1 hour and 7 minutes).

\smallskip\noindent
\textbf{Datasets.} Our datasets comprise \textit{real-world}, \textit{realistic} and \textit{synthetic} data. Our \textbf{Company ownership} experiments have been run over a real-world dataset (RW1721) extracted from the Italian Companies KG~\citep{DBLP:conf/ruleml/BellomariniBCGL20}, in a slice comprising the (continuous) ownership edges from 2017-21. The realistic datasets (MN7-MN28) represent company ownership graphs from 700 thousand to 2.8 million nodes generated in the likeness of a real-world dynamic structure, \Revision{with their ownership changes over 5 intervals and a high change rate, to test performance.} 
The datasets for the \textbf{Lehigh University Benchmark} are as described in the MeTeoR paper~\citep{DBLP:conf/aaai/WangHWG22}; the \textbf{Meteorological} experiments employ 3 datasets extracted from the~\cite{meteorologicalDataset} dataset at 5/50/500 stations.
For the \textbf{iTemporal} experiments, the synthetic datasets (S1-S10M) are generated with a random distribution over a given domain and include from 1K to 10M facts. 
The \textbf{Time Series} experiments were run over the NASDAQ Composite Index~\citep{nasdaq_omx_group_nasdaq_1971}, a daily time series from February 1971 to March 2023 comprising 13596 records. From this dataset, we extracted the last 1\% (NQ1), 10\% (NQ10), and 50\% (NQ50) so to have 4 total data sizes. 
Details about the datasets employed in the experiments can be found in the Appendix.

\subsection{Company Ownership Experiments}
\label{sub:company-ownership-experiments}

Similarly to Example~\ref{ex:running_example}, we ran experiments on company ownership changes over time.

\noindent
\textbf{Scenarios.} The experiment scenarios contain different elements: 1) \textit{Temporal}: diamond operator, recursion, and constraints on variables; 2) \textit{Negation}: recursion and stratified negation; 3) \textit{Aggregation}, temporal aggregations; 4) \textit{Diamond}: diamond operator and recursion; 5) \textit{Box}: box operator and recursion.
We ran each scenario on MN7-28 and RW1721; \textit{Box} and \textit{Diamond} were also run on MeTeoR 1.0.15 for comparison, as they do not include features not supported by MeTeoR.\footnote{Due to an incompatibility between data and the \textit{seminaive} evaluation in MeTeoR, which let the execution run indefinitely, we had to run the MeTeoR experiments in \textit{naive} mode.}
All scenarios were run with the \textit{always merge} strategy; \textit{Box} was tested also in the \textit{minimal} and \textit{earliest merge}. 

\smallskip\noindent
\textbf{Discussion.} The performance of \textit{Temporal}, \textit{Negation} and \textit{Aggregation} are shown in Figure~\ref{fig:experiments-full}a. They all show good scalability,
with the elapsed time increasing linearly over the dataset sizes. \textit{Temporal}, the most complex scenario, is the most expensive as well but with just over 182 secs for the biggest datasets, while \textit{Aggregation} and \textit{Negation} perform \Revision{two times} faster at 24-99 secs. Figure~\ref{fig:experiments-full}b shows the results for \textit{Box} and \textit{Diamond} in both Vadalog and MeTeoR\footnote{\Revision{Due to the difference in performance, the plot for MeTeoR is shown separately within the figure. Appendix, Figures~4-5 show a larger rendition of these curves.}}.
Vadalog performs in a similar linear-increase fashion to the other scenarios, \Revision{about 100 times} faster than MeTeoR, which exceeds the 4,000 secs threshold in the \RevisionTwo{MN20-28} datasets in both scenarios. 
Merge strategy-wise, we see that the \textit{always merge (am)} \textit{Box} is the most expensive, running at \RevisionTwo{180 secs for MN28}, while \textit{earliest merge (em)} \textit{Box} is
on average 8\% faster, and similar to those of \textit{Diamond} at around \RevisionTwo{41-170 secs.} 
The nature of the datasets, having many adjacent facts, explains it:
by merging strategically, fewer facts are sent down the pipeline, and fewer operations are needed while merging all the facts at one time may be superfluous. 
Figure~\ref{fig:experiments-full}c shows the performance on the real-world dataset. MeTeoR results are not shown as they all exceed the 
timeout. 
In Vadalog, \textit{Negation} and \textit{Temporal} are the fastest at 102 secs, while \textit{Aggregation} is the most expensive at 183 secs.
In MN7-28, \textit{Aggregation} performed well compared to \textit{Temporal} and \textit{Negation}, but in RW1721, it is the worst due to having \Revision{12} times more intervals, decreasing performance.
As for the \textit{merge strategies}, 
RW1721 shows similar performance to MN7-28,
with 
\textit{Box em}
the best performer at \RevisionTwo{131 secs} and 
\textit{Box am}
the worst at \RevisionTwo{145 secs.} 
The full result tables \Revision{and larger versions of all graphs} can be found in the Appendix.

\Revision{
\begin{figure*}
      \begin{subfigure}[b]{0.31\textwidth}
         \centering
            \includegraphics[width=\textwidth]{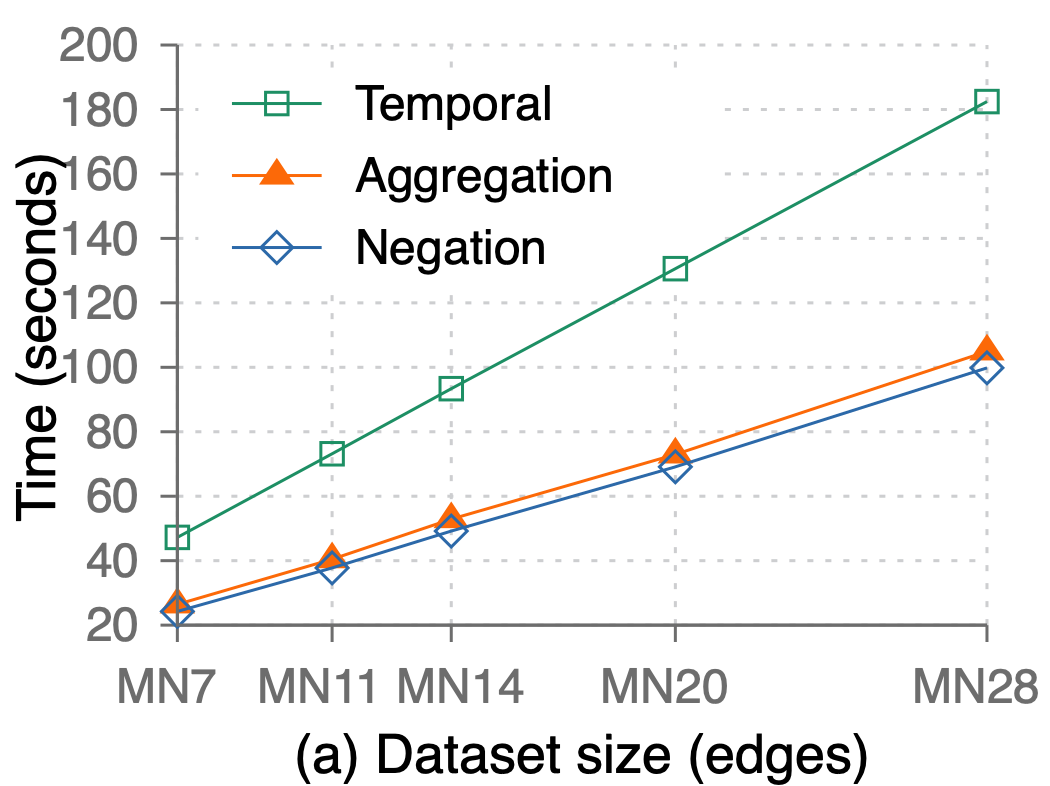}
         \label{fig:test1}
     \end{subfigure}
           \begin{subfigure}[b]{0.30\textwidth}
         \centering
       \includegraphics[width=\textwidth]{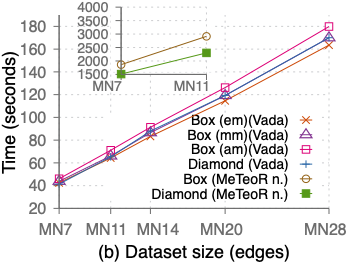}
         \label{fig:test2}
     \end{subfigure}
     \begin{subfigure}[b]{0.30\textwidth}
         \centering
        \includegraphics[width=\textwidth]{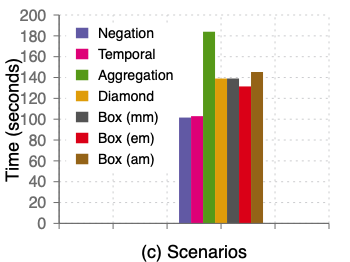}
         \label{fig:test3}
     \end{subfigure} 
      \begin{subfigure}[b]{0.30\textwidth}
         \centering
        \includegraphics[width=\textwidth]{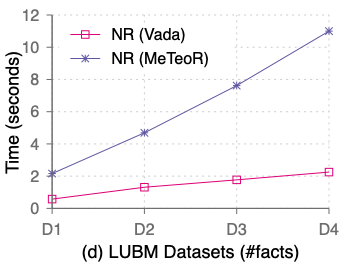}
         \label{fig:test8}
     \end{subfigure}
     \begin{subfigure}[b]{0.30\textwidth}
         \centering
        \includegraphics[width=\textwidth]{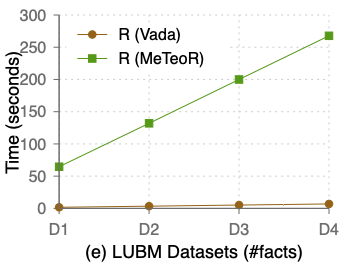}
         \label{fig:test9}
     \end{subfigure}   
      \begin{subfigure}[b]{0.30\textwidth}
         \centering
        \includegraphics[width=\textwidth]{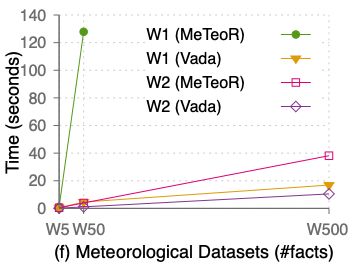}
         \label{fig:test6}
     \end{subfigure}    
      \begin{subfigure}[b]{0.30\textwidth}
         \centering 
         \includegraphics[width=\textwidth]{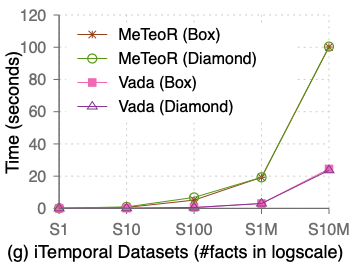}
         \label{fig:test4}
     \end{subfigure}
     \begin{subfigure}[b]{0.30\textwidth}
         \centering 
         \includegraphics[width=\textwidth]{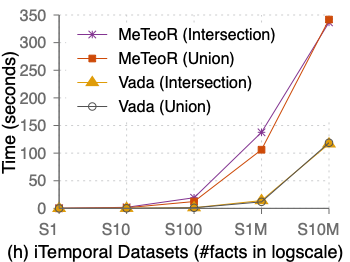}
         \label{fig:test5}
     \end{subfigure}
         \begin{subfigure}[b]{0.30\textwidth}
         \centering
        \includegraphics[width=\textwidth]{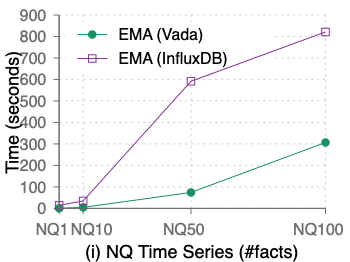}
         \label{fig:test7}
     \end{subfigure}
        \caption{\small{(a) \textit{Temporal}, \textit{Aggregation} and \textit{Negation} on MN7-MN28 over time; (b) \textit{Box} and \textit{Diamond} in Temporal Vadalog and MeTeoR; (c) RW dataset in all scenarios; (d) LUBM Non-Recursive experiment; (e) LUBM Recursive experiment; (f) Meteorological experiments; (g) \textit{iTemporal} Box and Diamond; (h) \textit{iTemporal} Union and Intersection; (i) Time Series.}}
        \label{fig:experiments-full}
\end{figure*}
}
\subsection{Lehigh University Benchmark Experiments (LUBM)}
\label{sub:lubm-experiments}
The LUBM benchmark for DatalogMTL~\citep{DBLP:conf/aaai/WangHWG22} in a modified form ~\citep{WALEGA2023100776} measures
the efficiency of forward-propagating rules in MeTeoR in the interval $[0,300]$ for four different dataset sizes (5, 10, 15, and 20 million facts) and two different programs (one recursive and one non-recursive program). 

\smallskip\noindent
\textbf{Discussion.} We used the published program and dataset from \cite{WALEGA2023100776} and repeated the experiments for MeTeoR \Revision{(\textit{seminaive} mode)} and our system. The results are presented in Figures~\ref{fig:experiments-full}\Revision{d-e} and show an overhead for MeTeoR of a factor of at least 3 for the non-recursive program (NR) and of a factor of at least 30 for the recursive ones (R). The main factor for better performance in Vadalog is the join, which in recursive programs especially shows the benefit of \Revision{indexing} already-seen atoms.

\subsection{Meteorological Experiments}
\label{sub:meteo-experiments}

We adapted the rules of the experiments on meteorological data~\citep{meteorologicalDataset} in~\cite{DBLP:conf/aaai/WangHWG22}
The programs W1 and W2 are run over a filtered version of this dataset: W1 regarding high temperatures, and W2 regarding heavy winds. 
For this reason, we 
%We
have to consider that while the original dataset is the same, W1 works with approximately \Revision{10 times} the facts of W2. \Revision{MeTeoR runs in \textit{seminaive} mode.}

\smallskip\noindent
\textbf{Discussion.} The results are shown in Figure~\ref{fig:experiments-full}f. Both W1 and W2 perform well in Vadalog, with Vadalog being \Revision{28 times} faster than MeTeoR for W1 at the W50 size (MeTeoR exceeds the time out for W500), and around \Revision{4 times} faster for W2.  

\subsection{iTemporal Experiments}
\label{sub:itemporal-experiments}
Considering the previous experiments, we observe a substantial speedup of the Temporal Vadalog System with respect to MeTeoR \Revision{(\textit{seminaive} mode)}. To confirm these measurements, we used \textit{iTemporal}~\citep{BellomariniNS22}, a generator for DatalogMTL, to generate specific experiments with different data sizes targeting the main temporal operations (that is, the application of temporal operators, joins, and unions). 

\smallskip\noindent
\textbf{Discussion.} In Figures~\ref{fig:experiments-full}\Revision{g-h} we present the results, which show that our system outperforms MeTeoR with a factor of 3 to 4 (depending on the benchmark) for $10M$ facts and \Revision{reinforces} the observations of the previous benchmarks \Revision{(e.g., it scales linearly)}.

\subsection{Time Series Experiments}
\label{sub:timeseries-experiments}

We executed the computation of the \textit{exponential moving average} (EMA) on Vadalog and the Time Series Database InfluxDB \Revision{(Version 2.6.1)} over the NASDAQ Composite Index~\citep{nasdaq_omx_group_nasdaq_1971}. \Revision{InfluxDB is a NoSQL time series database with a distinct data model optimized for time series operations.~\citep{InfluxDBKeyConcepts}}
For a fair comparison with the streaming-based approach of Vadalog, we compare the total time of loading and querying the data (end-to-end process).

\smallskip\noindent\textbf{Discussion.} Results are shown in Figure~\ref{fig:experiments-full}\Revision{i.} Our system outperforms InfluxDB by 2.5x for the largest dataset, considering the end-to-end measurement. We note that 
the experiment's aim is not to compare against the performance of InfluxDB, but to show that our system can achieve reasonable reasoning times in an end-to-end manner. 

\section{Related Work}
\label{sec:related-work}

The 1980s see the first proposals for a temporal version of Datalog with Datalog\textsubscript{1S}~\citep{Chomicki90,ChomickiI88,Chomicki89}, using successor functions, and Datalog extensions with temporal logic operators from Linear Temporal Logic (LTL) \Revision{with Templog~\citep{AbadiM89},} \Revision{both with further work from Baudinet~\citeyearpar{baudinet1992simple,DBLP:books/bc/tanselCGSS93/BaudinetCW93,DBLP:journals/iandc/Baudinet95}} and Computation Tree Logic (CTL) \Revision{with Datalog LITE~\citep{GoGrVe2002}.} \Revision{Further work on constraint databases include:~\cite{DBLP:journals/constraints/BrodskyJM97}, \cite{DBLP:conf/pods/MaherS96}, \cite{DBLP:books/sp/Revesz02}.}
\Revision{Interest in the application of logic programming over the dense timeline focused on Metric Temporal Logic~\citep{ALUR199335,koymans_specifying_1990} in the work of ~\cite{DBLP:conf/slp/Brzoska95,DBLP:journals/tcs/Brzoska98}. More recently, MTL has been considered as a formalism that can provide an expressive temporal extension to Datalog with DatalogMTL in the work of~\cite{BrKR18,AAAI1714881}, who also presented a practical implementation through SQL rewriting, although non-recursive.}
Theory for continuous semantics DatalogMTL~\citep{DBLP:conf/aaai/CucalaWGK21,WCKK19,DBLP:conf/kr/WalegaGKK20,DBLP:conf/ijcai/WalegaGKK20,WaKC19}
%,WCKG21,WCKK19,DBLP:conf/kr/WalegaGKK20,DBLP:conf/ijcai/WalegaGKK20,WaKC19,WalegaZG23} 
as well as for pointwise~\citep{DBLP:conf/dlog/KikotRWZ18} has been studied extensively in recent years, and most recently with MeTeoR~\citep{DBLP:conf/aaai/WangHWG22,Walega23}, a reasoner employing a combination of materialization and automata-based reasoning, and that we have referred for our comparative experiments. 

\Revision{Another comparable system, the Dyna language solver, proposed by~\cite{VieiraFFKE17} employs fixed point evaluation but computation may fail to terminate, while Vadalog employs the isomorphic chase with termination strategies, hence a program will always terminate; 
the Dyna solver uses forward or backward chaining depending on the strategy chosen by its reinforcement learning optimizer, whereas Vadalog uses a pull-based approach for evaluation, more efficient in resource usage for query-answering.}

The non-temporal version of the Vadalog System~\citep{BeBS20} has been proposed as a system to reason over KGs~\citep{BeSV20} with rules expressed in Warded Datalog$^\pm$~\citep{BeSG18}.
Other reasoners of similar expressive power include PDQ~\citep{BeLT15}, Llunatic~\citep{GMPS13}, Graal~\citep{BLMR15}, DLV~\citep{LPFE06}, and RDFox~\citep{NPMH15}. 

\section{Conclusion}
\label{sec:conclusions}

In this paper, we introduced a new framework and architecture for reasoning with DatalogMTL, highlighting its efficiency and ability to handle complex tasks, like the reasoning over time series, through multiple performance evaluations. We exceeded the capabilities of existing DatalogMTL reasoning tools. Moving forward, we aim to enhance the reasoner and explore other fragments of DatalogMTL.

\bibliographystyle{tlplike}
\bibliography{b}

\end{document}

\lefttitle{Bellomarini, L., Blasi, L., Nissl, M., Sallinger, E.}

\jnlPage{1}{17}
\jnlDoiYr{2021}
\doival{10.1017/xxxxx}

\title[Theory and Practice of Logic Programming]{The Temporal Vadalog System: \\ Appendix}

\begin{authgrp}
\author{\gn{Luigi} \sn{Bellomarini} }
\affiliation{Bank of Italy, Italy}
\author{\gn{Livia} \sn{Blasi} }
\affiliation{Bank of Italy, Italy} \affiliation{TU Wien, Austria}
\author{\gn{Markus} \sn{Nissl}}
\affiliation{TU Wien, Austria}
\author{\gn{Emanuel} \sn{Sallinger}}
\affiliation{TU Wien, Austria} \affiliation{University of Oxford, UK}

\end{authgrp}

\history{\sub{31-03-2023}}

\maketitle

\section{Merging Strategies: Blocking Example}

In the following section, we will present an example of a run of the \textit{Blocking} algorithm presented in the main paper, in Algorithm~2.

Consider the following program:

\begin{verbatim}
b(X,Y):- sourceNode(X,Y).
a(X,Y):- b(Y,X).
b(X,Y):- a(X,Y).
outputNode(X):- a(X,Y).
\end{verbatim}

And consider that the filter for \textit{a} includes the placement of the MergeNode.

Our database, represented in the program by \textit{sourceNode}, includes the following facts: 

\begin{verbatim}
sourceNode(1,2)@[1,3].
sourceNode(1,2)@[5,6].
sourceNode(2,1)@[0,1].
sourceNode(1,3)@[9,10].
\end{verbatim}

Initially, MergeNode is set to:

\begin{verbatim}
currentPosition: 0
mergeStructure: empty (length: 0)
\end{verbatim}

\subsection{First Next Call}
The computation starts when our sink (\textsc{outputNode}) calls \textsc{next()} from its upstream filter (\textsc{A}).

We enter the MergeNode and on Line~4 we set:
\begin{verbatim}
changed: false
\end{verbatim}

Lines~5-8: 
\texttt{currentPosition} is equal to \texttt{mergeStructure.length}, so we call the 
upstream filter (\textsc{B}) repeatedly, until there are no more facts to pull. 
The current content of \texttt{mergeStructure} becomes thus:

\begin{verbatim}
mergeStructure (length: 3): 
a(2,1): [1,3], [5,6]
a(1,2): [0,1]
a(3,1): [9,10]

changed: true
\end{verbatim}
While \texttt{changed} is set to \textit{true} (Line~8).

Line~9-10: as \texttt{changed} is true, \texttt{currentPosition} is set to 0.

Line~11: The cursor for the retrieval of an entry is set to the position in \texttt{mergeStructure} correspondent to \texttt{currentPosition}. 

Line~12-13: \texttt{currentPosition} is incremented and becomes 1 and we return \textit{true}.

\subsection{Second Next Call}

\textsc{outputNode} calls \textsc{next} from \textsc{A} for the second time.

Since \texttt{currentPosition < mergeStructure.length} we only run Lines~11-12:
\begin{verbatim}
set(mergeStructure(1))
currentPosition: 2
\end{verbatim}

\subsection{Third Next Call}
Third \textsc{next} call:
we run Lines~11-12:
\begin{verbatim}
set(mergeStructure(2))
currentPosition: 3
\end{verbatim}

\subsection{Fourth Next Call}
Fourth \textsc{next} call:
\texttt{currentPosition >= mergeStructure.length} (3 $\geq$ 3) so we check for new facts from the upstream filter.

Since in the meantime not only \textsc{outputNode}, but filter \textsc{B} too pulled from \textsc{A} and transformed them as such:

\begin{verbatim}
b(2,1)@[[1,3],[5,6]]
b(1,2)@[[0,1]]
b(3,1)@[[9,10]]
\end{verbatim}

The \textsc{next} function can pull new facts from \textsc{B} and hence the \texttt{mergeStructure} becomes:

\begin{verbatim}
mergeStructure: 
a(2,1): [0,3], [5,6]
a(1,2): [0,3], [5,6]
a(3,1): [9,10]
a(1,3): [9,10]

changed: true
\end{verbatim}

And here we follow again:

Line~9-10: as \texttt{changed} is \textit{true}, \texttt{currentPosition} is set to 0.

Line~11: The cursor for the retrieval of an entry is set to the position in \texttt{mergeStructure} correspondent to \texttt{currentPosition}. 

Line~12-13: \texttt{currentPosition} is incremented and becomes 1 and we return \textit{true}.

\subsection{Subsequent Next Calls}

The four facts in \texttt{mergeStructure} are pulled from the fourth to the eight \textsc{next} calls.

\medskip\noindent
At the end of the ninth call of \textsc{next}, since \textsc{B} cannot produce more new facts from the facts pulled from \textsc{A}, we finally return \textit{false}, and the procedure ends.

\section{Experiments: additional material}

This section includes the supplementary material of the experiments on Temporal Vadalog, mostly comprised of the tables covering the dataset information and the results.

\subsection{Datasets}

Figure~\ref{fig:dataset-info} shows the dataset info for the Company Ownership Experiments, with \textit{realistic} datasets MN7-28 and \textit{real-world} dataset RW1721. 

\RevisionTwo{The \textit{realistic} datasets MN7-28 have been generated as scale-free networks per the Barabasi-Albert model~\citep{Barabasi}.}

\medskip
In Figure~\ref{fig:dataset-2-info}, we present the dataset information for the \textit{iTemporal}~\citep{BellomariniNS22} experiments, for the Lehigh University Benchmark Experiments, for the Time Series Experiments and for the Meteorological Experiments.

The Time Series dataset is the NASDAQ Composite Index~\citep{nasdaq_omx_group_nasdaq_1971}.

The Meteorological dataset~\citep{meteorologicalDataset} has been used in two programs that look at different aspects of the data collected by the meteorological stations: W1 covers high temperature, while W2 covers heavy winds. The data about the subset of datasets has been reported in the table as ``\{Dataset\} filter \{Program\}'' (e.g. ``W5 subset for W1'' is the subset of W5 for the program W1).

\begin{figure}
\caption{\small{Details about the size and features of the datasets employed in both the realistic and real-world Company Ownership Experiments.} }

\small{
 {\tablefont\begin{tabularx}{\textwidth}{@{\extracolsep{\fill}}llll}
\topline
Name   & \multicolumn{1}{l}{\#Edges} & \multicolumn{1}{l}{\#Nodes} & Intervals           \midline
MN7   & 2,795,807              & 700,000                & 5 intervals        \\
MN11   & 4,345,074              & 1,100,000              & 5 intervals        \\
MN14   & 5,537,868              & 1,400,009              & 5 intervals        \\
MN20   & 7,781,459              & 2,000,000              & 5 intervals        \\
MN28   & 10,901,234             & 2,800,000              & 5 intervals        \\
RW1721 & 13,588,467             & 8,162,205              & m. 2017-21   \botline
\end{tabularx}
}}

\label{fig:dataset-info}
\end{figure}

\begin{figure}
\caption{\small{Details about the iTemporal, LUBM, Meteorological and Time Series datasets.} }

 {\tablefont\begin{tabularx}{\textwidth}{@{\extracolsep{\fill}}@{}llll@{}}
\topline
\textbf{Experiment}     & \textbf{Dataset Name} & \textbf{Type} & \textbf{\# Facts} \midline
\textbf{iTemporal}      & S001                  & synthetic     & 10                \\
\textbf{}               & S01                   & synthetic     & 100               \\
\textbf{}               & S1                    & synthetic     & 1,000             \\
\textbf{}               & S10                   & synthetic     & 10,000            \\
\textbf{}               & S100                  & synthetic     & 100,000           \\
\textbf{}               & S1M                   & synthetic     & 1,000,000         \\
\textbf{}               & S10M                  & synthetic     & 10,000,000        \midline
\textbf{LUBM}           & D1                    & synthetic     & 5,000,000         \\
\textbf{}               & D2                    & synthetic     & 10,000,000        \\
\textbf{}               & D3                    & synthetic     & 15,000,000        \\
\textbf{}               & D4                    & synthetic     & 20,000,000        \midline
\textbf{Time Series}    & NQ1                   & real-world    & 135               \\
\textbf{}               & NQ10                  & real-world    & 1,350             \\
\textbf{}               & NQ50                  & real-world    & 6,798             \\
\textbf{}               & NQ100                 & real-world    & 13,596            \midline
\textbf{Meteorological} & W5                    & real-world    & 177,145           \\
\textbf{}               & W50                   & real-world    & 1,771,450         \\
\textbf{}               & W500                  & real-world    & 17,714,501        \midline
\textbf{}               & W5 subset for W1          & real-world    & 97,943            \\
\textbf{}               & W50 subset for W1         & real-world    & 974,940           \\
\textbf{}               & W500 subset for W1        & real-world    & 9,744,923         \midline
\textbf{}               & W5 subset for W2          & real-world    & 9,811             \\
\textbf{}               & W50 subset for W2         & real-world    & 93,631            \\
\textbf{}               & W500 subset for W2        & real-world    & 931,831          

 \botline
\end{tabularx}}

\label{fig:dataset-2-info}
\end{figure}

\pagebreak

\section{Company Ownership Experiments}

\subsection{Rules}

\subsubsection{Scenario: Temporal}

\begin{verbatim}
prevOwn(X,Y,S):- <+>[0,1)companyOwn(X,Y,S).
ownershipChange(X,Y,Diff):- watchCompany(Y), 
        <->[0,1) companyOwn(X,Y,S), prevOwn(X,Y,S1), S<>S1, Diff=S-S1.
watchCompany(Z) :- ownershipChange(X,Y,Diff), connected(Y,Z).

connected(Y,Z):- companyOwn(X,Y,S), companyOwn(X,Z,S), Y != Z.
\end{verbatim}

\subsubsection{Scenario: Negation}

\begin{verbatim}
succOwn(X,Y,S):- <+>[0,1)companyOwn(X,Y,S).
ownershipLoss(X, Y):- <->[0,1) companyOwn(X,Y,S), not succOwn(X,Y,S1), 
        watchCompany(Y).
watchCompany(Z) :- ownershipLoss(X,Y), connected(Y,Z).

connected(Y,Z):- companyOwn(X,Y,S), companyOwn(X,Z,S), Y != Z.
\end{verbatim}

\subsubsection{Scenario: Aggregation}

\begin{verbatim}
ownershipTotal(Y, Z) :- companyOwn(X,Y,S), Z=msum(S).
\end{verbatim}

\subsubsection{Scenario: Box}

\paragraph{Temporal Vadalog}

\begin{verbatim}
ownershipChange(X,Y):- watchCompany(Y), [-][0,3) companyOwn(X,Y,S).
watchCompany(Z) :- ownershipChange(X,Y), connected(Y,Z).

connected(Y,Z):- companyOwn(X,Y,S), companyOwn(X,Z,S).    
\end{verbatim}

Run with different \textit{merge strategies}: \textit{always merge (am)}, \textit{earliest merge (em)}, \textit{minimal merge (mm)}. A detail of the experiment for this can be seen in Figure~\ref{fig:inverse}.

\paragraph{MeTeoR}

\begin{Verbatim}[commandchars=\\\{\}]
OwnershipChange(X,Y):-WatchCompany(Y),\RevisionEnvTwo{Boxminus[0,3)CompanyOwn(X,Y,S)}
WatchCompany(Z) :-OwnershipChange(X,Y),Connected(Y,Z)
\RevisionEnvTwo{Connected(Y,Z):- CompanyOwn(X,Y,S), CompanyOwn(X,Z,S)}
\end{Verbatim}

\subsubsection{Scenario: Diamond}

\paragraph{Temporal Vadalog}

\begin{verbatim}
ownershipChange(X,Y):- watchCompany(Y), <->[0,1) companyOwn(X,Y,S).
watchCompany(Z) :- ownershipChange(X,Y), connected(Y,Z).
connected(Y,Z):- companyOwn(X,Y,S), companyOwn(X,Z,S).    
\end{verbatim}

\paragraph{MeTeoR}

\begin{Verbatim}[commandchars=\\\{\}]
OwnershipChange(X,Y):-WatchCompany(Y),Diamondminus[0,1)\RevisionEnvTwo{CompanyOwn(X,Y,S)}
WatchCompany(Z):-OwnershipChange(X,Y),Connected(Y,Z)
\RevisionEnvTwo{Connected(Y,Z):-CompanyOwn(X,Y,S),CompanyOwn(X,Z,S)}    
\end{Verbatim}

\subsubsection{Results}

\begin{figure}
    \centering
    \includegraphics[width=0.95\textwidth]{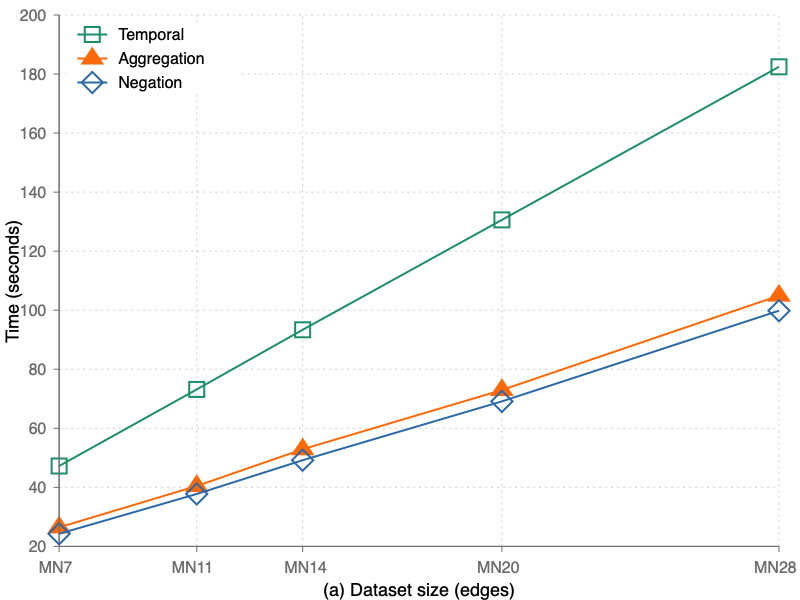}
    \caption{Larger version of the Company Ownership experiments over the \textit{Negation}, \textit{Aggregation} and \textit{Temporal} scenarios. }
    \label{fig:scalability}
\end{figure}

\begin{figure}
    \centering
    \includegraphics[width=0.95\textwidth]{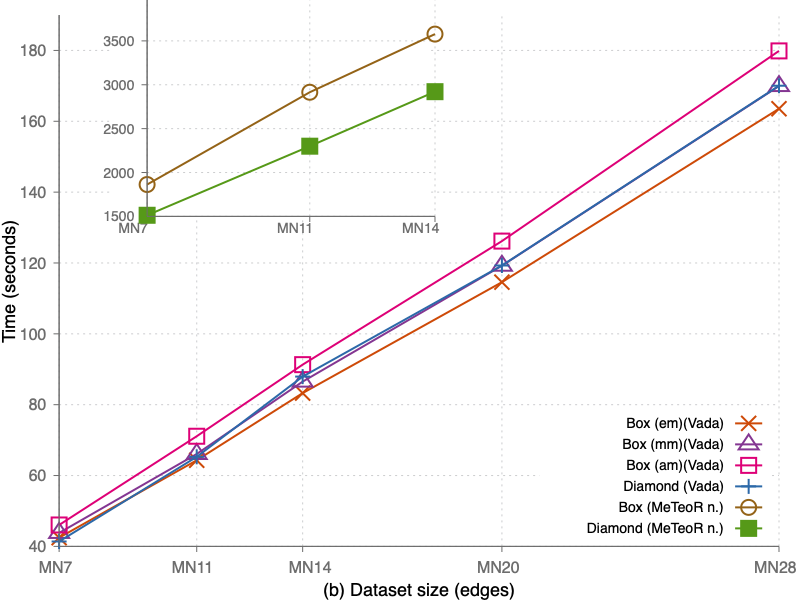}
    \caption{Larger version of the detail about the results for the \textit{Box} (merge strategies: \textit{always merge (am)}, \textit{earliest merge (em)} and \textit{minimal merge (mm)}) and \textit{Diamond} scenario in Vadalog; the results in MeTeoR are shown in the mini plot.}
    \label{fig:inverse}
\end{figure}

\begin{figure}
    \centering
    \includegraphics[width=0.95\textwidth]{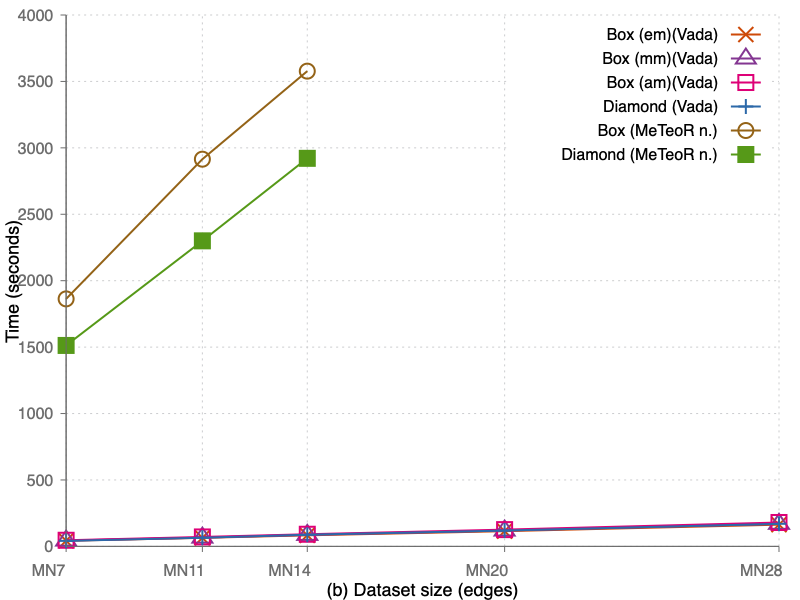}
    \caption{\textit{Box} (merge strategies: \textit{always merge (am)}, \textit{earliest merge (em)} and \textit{minimal merge (mm)}) and \textit{Diamond} scenario in Vadalog and MeTeoR.}
    \label{fig:vada-meteor}
\end{figure}

\begin{figure}
    \centering
    \includegraphics[width=0.95\textwidth]{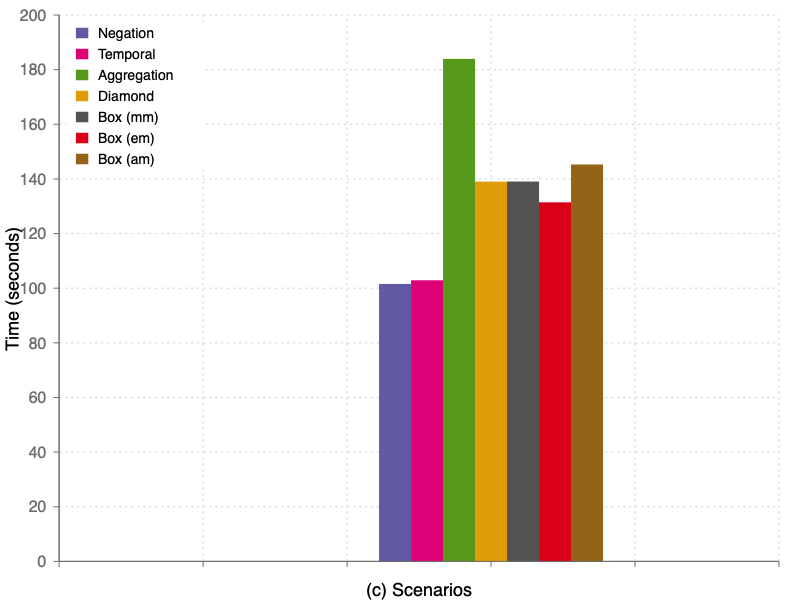}
    \caption{Larger version of the Company Ownership experiments on the Real World dataset. }
    \label{fig:real-world-graph}
\end{figure}

Figure~\ref{fig:table-results-co} shows the results in seconds for the execution elapsed time of the Company Ownership Experiments, all 5 scenarios, in Temporal Vadalog, while Figure~\ref{fig:table-meteor-results-co} shows the same results for the applicable scenarios in MeTeoR~\citep{DBLP:conf/aaai/WangHWG22}.
Figures~\ref{fig:scalability}, \ref{fig:inverse} and \ref{fig:real-world-graph} are larger versions of the graphs shown in the main contribution, while Figure~\ref{fig:vada-meteor} shows the Company Ownership scenarios in both Vadalog and MeTeoR in a single graph as to visually show the difference in performance.

\begin{figure}[t]
\centering
\caption{The results of the Company Ownership Experiments in Temporal Vadalog. The results are expressed in seconds. }
\label{fig:table-results-co} {\tablefont\begin{tabularx}{\textwidth}{@{\extracolsep{\fill}}llllllll}
\topline

 & \multicolumn{7}{l}{\textbf{Temporal Vadalog}}                                                                   
 \midline
   \textbf{Data} & Temporal   & Negation   & Aggregation & Diamond    & Box (mm) & Box (em) & Box (am)
    \midline
MN7 & 47.23    & 24.26    & 26.44       & 41.38   & 43.80    & 42.45    & 46.04\\
MN11 & 73.17    & 37.75    & 40.45       & 65.28   & 66.08    & 64.31    & 71.09    \\
MN14 & 93.37    & 49.19    & 52.90       & 87.93   & 86.51    & 83.27    & 91.30    \\
MN20 & 130.60   & 69.12    & 72.97       & 119.29  & 119.23   & 114.59   & 126.18   \\
MN28 & 182.45   & 99.83    & 104.94      & 170.04  & 169.95   & 163.54   & 179.87   \\
RW1721   & 102.89   & 101.55   & 183.94      & 139.01  & 139.04   & 131.43   & 145.28  
        \botline
\end{tabularx}}
\end{figure}

\begin{figure}[t]
%\begin{tabular}{llllllll}
\centering
\caption{The results of the Company Ownership Experiments in MeTeoR. The results are expressed in seconds. }
\label{fig:table-meteor-results-co} {\tablefont\begin{tabularx}{0.5\textwidth}{@{\extracolsep{\fill}}lll}

\topline
     \multicolumn{3}{l}{\textbf{MeTeoR ``naive'' mode 1.0.15}} \midline
    \textbf{Data} & Diamond          & Box              \midline
MN7 & \RevisionEnvTwo{1512.61}          & \RevisionEnvTwo{1863.02}          \\
MN11 & \RevisionEnvTwo{2299.96}          & \RevisionEnvTwo{2914.74}          \\
MN14 & \RevisionEnvTwo{2921.11}         & \RevisionEnvTwo{3577.46}         \\
MN20 & \RevisionEnvTwo{time out}         & \RevisionEnvTwo{time out}         \\
MN28 & time out         & time out         \\
RW1721   & \RevisionEnvTwo{time out}         & \RevisionEnvTwo{time out}            
\botline
\end{tabularx}}
\end{figure}

\section{iTemporal Experiments}

\subsection{Benchmark Generation parameters}
The benchmark generator with the version with the git hash \#105fa48c148ae3f63998de1f6b80e9d63a903428 was used with the following parameters:

\begin{verbatim}
    {
  "properties": {
	"nodes": 4,
	"inputNodes": 2,
	"outputNodes": 1,
	"multiEdgeRules": 0.3,
	"recursiveRules": 0.1,
	"recursiveComplexity": 0.1,
	"coreMultiEdgeRules": 1,
	"multiEdgeTemporalRules": 0,
	"coreSingleEdgeRules": 1,
	"singleEdgeTemporalRules": 0,
	"aggregationRules": 0,
	"linearRules": 1,
	"intersectionRules": 1,
	"unionRules": 0,
	"diamondMinusRules": 0.4,
	"diamondPlusRules": 0.4,
	"boxMinusRules": 0.1,
	"boxPlusRules": 0.1,
	"sinceRules": 1,
	"untilRules": 0,
	"spanningTemporalAggregationRules": 0.3,
	"movingWindowTemporalAggregationRules": 0.3,
	"instantaneousTemporalAggregationRules": 0.4,
	"averageOutputArity": 2,
	"varianceOutputArity": 0,
	"averageNumberOfContributorTerms": 1,
	"varianceNumberOfContributorTerms": 1,
	"averageNumberOfGroupByTerms": 2,
	"varianceNumberOfGroupByTerms": 1,
	"averageNumberOfOverlappingJoinTerms": 1,
	"varianceNumberOfOverlappingJoinTerms": 1,
	"temporalFactor": 1000,
	"averageNumberOfTemporalUnitsT1": 0,
	"varianceNumberOfTemporalUnitsT1": 10,
	"averageNumberOfTemporalUnitsT2": 100,
	"varianceNumberOfTemporalUnitsT2": 10,
	"temporalMaxPrecision": 0,
	"cardinalityTermDomain": 1000,
	"averageAmountOfGeneratedOutputs": 10,
	"varianceAmountOfGeneratedOutputs": 0,
	"outputTimestampStart": 1577836800000,
	"outputTimestampEnd": 1577836904000,
	"averageOutputIntervalDuration": 200,
	"varianceOutputIntervalDuration": 100,
	"averageAggregationSelectivity": 0.2,
	"varianceAggregationSelectivity": 0.02,
	"unionInclusionPercentage": 0.6,
	"temporalInclusionPercentage": 0.6,
	"path": "xxxx",
	"generateTimePoints": false,
	"averageAggregationBucket": 1,
	"varianceAggregationBucket": 0,
	"percentageViaContributor": 0.7,
	"outputCsvHeader": true,
	"outputQuestDB": false
  },
  "graphInternal": {
	"nodes": [
  	{
    	"name": "g50",
    	"type": "Input",
    	"minArity": 2,
    	"maxArity": 2,
    	"isCyclic": false,
    	"sccId": 2
  	},
  	{
    	"name": "g51",
    	"type": "Input",
    	"minArity": 2,
    	"maxArity": 2,
    	"isCyclic": false,
    	"sccId": 3
  	},
  	{
    	"name": "g53",
    	"type": "Output",
    	"minArity": 2,
    	"maxArity": 2,
    	"isCyclic": false,
    	"sccId": 1
  	}
	],
	"edges": [
  	{
    	"from": "g50",
    	"to": "g53",
    	"type": "UnionEdge",
    	"isCyclic": false,
    	"uniqueId": "g67",
    	"termOrderShuffleAllowed": true,
    	"termOrder": [
      	0,
      	1
    	],
    	"overlappingTerms": 2,
    	"nonOverlappingTerms": 0,
    	"isLeftEdge": false,
    	"aggregationType": "Unknown",
    	"numberOfGroupingTerms": -1,
    	"numberOfContributors": -1,
    	"t1": -1,
    	"t2": -1,
    	"unit": "Unknown"
  	},
  	{
    	"from": "g51",
    	"to": "g53",
    	"type": "UnionEdge",
    	"isCyclic": false,
    	"uniqueId": "g68",
    	"termOrderShuffleAllowed": true,
    	"termOrder": [
      	0,
      	1
    	],
    	"overlappingTerms": 2,
    	"nonOverlappingTerms": 0,
    	"isLeftEdge": false,
    	"aggregationType": "Unknown",
    	"numberOfGroupingTerms": -1,
    	"numberOfContributors": -1,
    	"t1": -1,
    	"t2": -1,
    	"unit": "Unknown"
  	}
	]
  }

\end{verbatim}

\subsection{Rules}

\subsubsection{iTemporal Box}

\begin{verbatim}
g18(N0) :- [-][50000.0,247000.0] g17(N0).
\end{verbatim}

\subsubsection{iTemporal Diamond}

\begin{verbatim}
g18(N0) :- <->[0.0,11.0] g17(N0).
\end{verbatim}

\subsubsection{iTemporal Union}

\begin{verbatim}
g53(N0,N1) :- g50(N0,N1).
g53(N0,N1) :- g51(N0,N1).
\end{verbatim}

\subsubsection{iTemporal Intersection}

\begin{verbatim}
g53(N0,N1) :- g50(N0,N1), g51(N0).
\end{verbatim}

\subsection{Results}

\begin{figure}
    \centering
    \includegraphics[width=0.95\textwidth]{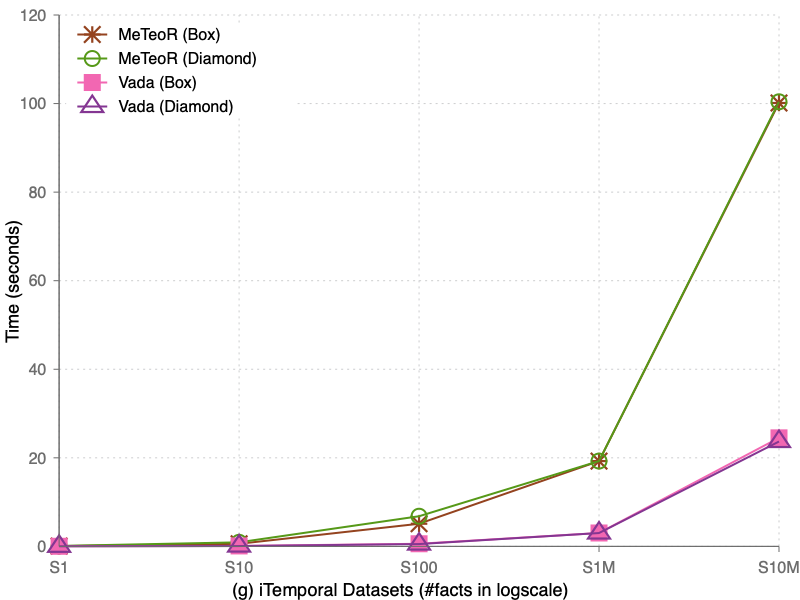}
    \caption{Larger version of \textit{iTemporal} experiments for Box and Diamond.}
    \label{fig:itemporal-bd}
\end{figure}

\begin{figure}
    \centering
    \includegraphics[width=0.95\textwidth]{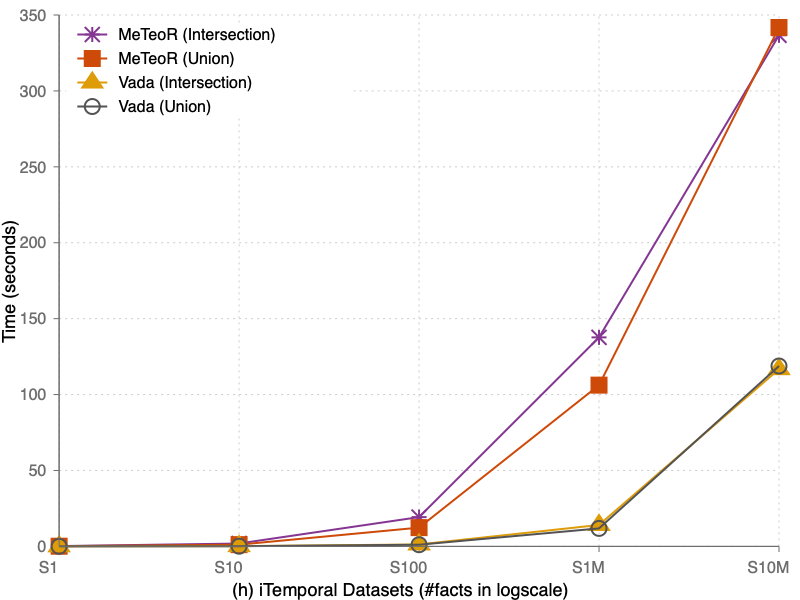}
    \caption{Larger version of \textit{iTemporal} experiments for Union and Intersection. }
    \label{fig:itemporal-iu}
\end{figure}

Figure~\ref{fig:table-itemporal-results} shows the results for the \textit{iTemporal} experiments in Temporal Vadalog and MeTeoR 1.0.15. The latter has been run in the mode \textit{seminaive}. Figures~\ref{fig:itemporal-bd}-~\ref{fig:itemporal-iu} are larger versions of the graphs shown in the main contribution.

\begin{figure}[t]
%\begin{tabular}{llllllll}
\centering
\caption{The results of the iTemporal Experiments. The results are expressed in seconds. }
\label{fig:table-itemporal-results} {\tablefont\begin{tabularx}{\textwidth}{@{\extracolsep{\fill}}lllllllll}

\topline

Program:        & \multicolumn{2}{l}{\textbf{Box}}   & \multicolumn{2}{l}{\textbf{Diamond}} & \multicolumn{2}{l}{\textbf{Intersection}} & \multicolumn{2}{l}{\textbf{Union}} \midline
\textbf{Dataset} & \textbf{Vadalog} & \textbf{MeTeoR} & \textbf{Vadalog}  & \textbf{MeTeoR}  & \textbf{Vadalog}     & \textbf{MeTeoR}    & \textbf{Vadalog} & \textbf{MeTeoR} \midline
S001              & 0.01567          & 0.00887         & 0.00767           & 0.00519          & 0.01200              & 0.01450            & 0.01367          & 0.00536         \\
S01             & 0.01333          & 0.00994         & 0.00867           & 0.01348          & 0.01333              & 0.02291            & 0.01233          & 0.01410         \\
S1              & 0.02233          & 0.06118         & 0.01767           & 0.09918          & 0.04133              & 0.19526            & 0.02100          & 0.11512         \\
S10             & 0.10133          & 0.58417         & 0.07367           & 0.92113          & 0.16167              & 1.90043            & 0.10700          & 1.14543         \\
S100            & 0.59067          & 5.11834         & 0.52200           & 6.80771          & 1.27567              & 19.28032           & 1.09067          & 12.35570        \\
S1M              & 3.02500          & 19.26079        & 3.02000           & 19.26737         & 14.11867             & 137.66972          & 11.83767         & 106.20178       \\
S10M             & 24.50633         & 100.12875       & 23.68333          & 100.40331        & 116.73533            & 336.64323          & 118.79900        & 341.77622               
\botline
\end{tabularx}}
\end{figure}

\section{Lehigh University Benchmark (LUBM) Experiments}

\subsection{Rules}
For the MeTeoR rules, we refer to the Github Repository for MeTeoR, in the AAAI22 Experiments folder~\footnote[1]{https://github.com/wdimmy/MeTeoR/tree/main/experiments/AAAI2022/}.

\subsubsection{Program: R (Recursive)}

\begin{verbatim}
a1FullProfessor2(X) :- a1FullProfessor(X).

a1ScientistCandidate(X) :- <->[1,1]a1doctoralDegreeFrom(X,Y).
a1Scientist(X) :- [-][1,1]a1ScientistCandidate(X).
a1Scientist(X) :-  <->[1,1]a1FullProfessor2(X).
a1FullProfessor2(X) :-  <->[1,1]a1Scientist(X).
\end{verbatim}

\subsubsection{Program: NR (Non-Recursive)}

\begin{verbatim}
a1AssistantProfessorCandidate(X) :- <->[1,1]a1Lecturer(X).
a1AssociateProfessorCandidate(X) :- a1doctoralDegreeFrom(X,Y),
        [-][1,1]a1University(Y).
a1AssociateProfessorCandidate(X) :- [-][1,1]a1AssistantProfessorCandidate(X).
\end{verbatim}

\subsection{Results}

\begin{figure}
    \centering
    \includegraphics[width=0.95\textwidth]{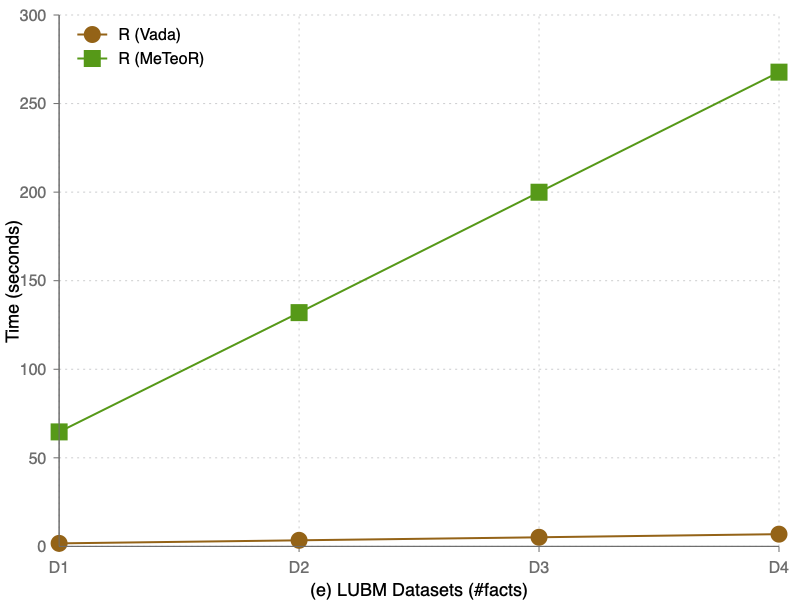}
    \caption{Larger version of LUBM experiments for the recursive program. }
    \label{fig:recursive-lubm}
\end{figure}

\begin{figure}
    \centering
    \includegraphics[width=0.95\textwidth]{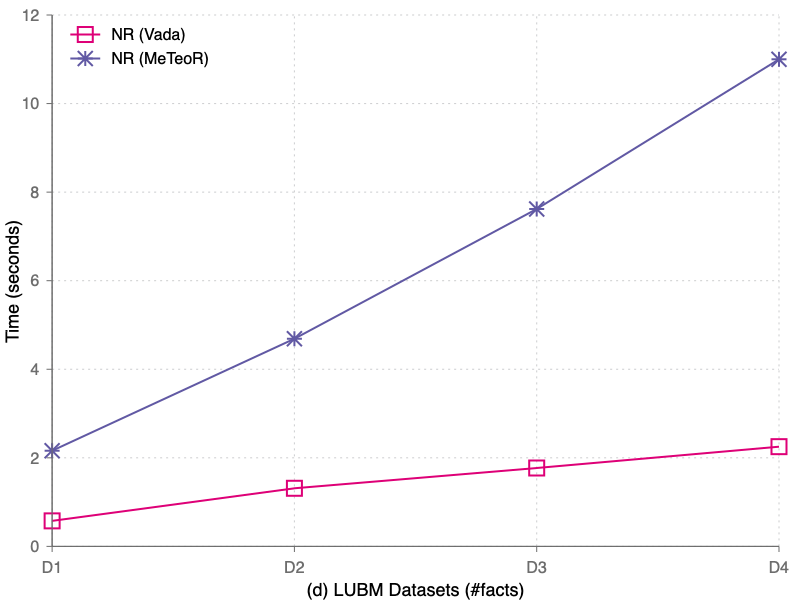}
    \caption{Larger version of LUBM experiments for the non-recursive program. }
    \label{fig:nonrecursive-lubm}
\end{figure}

Figure~\ref{fig:table-lubm-results} shows the results of the experiments over the LUBM datasets for the Recursive (R) and Non-Recursive (NR) programs, in Temporal Vadalog and MeTeoR (v. 1.0.15, mode \textit{seminaive}). Figures~\ref{fig:recursive-lubm}-~\ref{fig:nonrecursive-lubm} are larger versions of the graphs shown in the main contribution.

\begin{figure}[t]
%\begin{tabular}{llllllll}
\centering
\caption{The results of the LUBM Experiments in MeTeoR, expressed in seconds. }
\label{fig:table-lubm-results} {\tablefont\begin{tabularx}{0.8\textwidth}{@{\extracolsep{\fill}}lllll}

\topline
Program:         & \multicolumn{2}{l}{\textbf{NR}}    & \multicolumn{2}{l}{\textbf{R}}     \midline
\textbf{Dataset} & \textbf{Vadalog} & \textbf{MeTeoR} & \textbf{Vadalog} & \textbf{MeTeoR} \midline
D1               & 0.576            & 2.160           & 1.697            & 64.632          \\
D2               & 1.312            & 4.689           & 3.433            & 131.972         \\
D3               & 1.770            & 7.619           & 5.146            & 199.943         \\
D4               & 2.252            & 11.000          & 6.897            & 267.739        
\botline
\end{tabularx}}
\end{figure}

\section{Meteorological Experiments}

\subsection{Rules}
The Temporal Vadalog rules used for the programs W1 and W2 are in the following paragraphs.
For the MeTeoR rules, we refer to the Github Repository for MeTeoR, in the AAAI22 Experiments folder\RevisionTwo{~\footnote[1]{\RevisionTwo{https://github.com/wdimmy/MeTeoR/tree/main/experiments/AAAI2022/}}}.

\subsubsection{Program: W1 (Heat)}

\begin{verbatim}
excessiveHeat(X) :- <+>[0,1] temp1(X).
temp1(X) :- <->[0,1] tempAbove41a(X), [-][0,1]tempAbove24a(X).
heatAffectedState(X) :- excessiveHeat(Y),locatedInState(Y,X).    
\end{verbatim}

\subsubsection{Program: W2 (HeavyWind)}

\begin{verbatim}
heavyWind(X) :- <+>[0,1] temp1(X).
temp1(X) :- [-][0,1] heavyWindForce(X).
heavyWindAffectedState(X) :- heavyWind(Y),locatedInState(Y,X).

\end{verbatim}

\subsection{Results}

\begin{figure}
    \centering
    \includegraphics[width=0.95\textwidth]{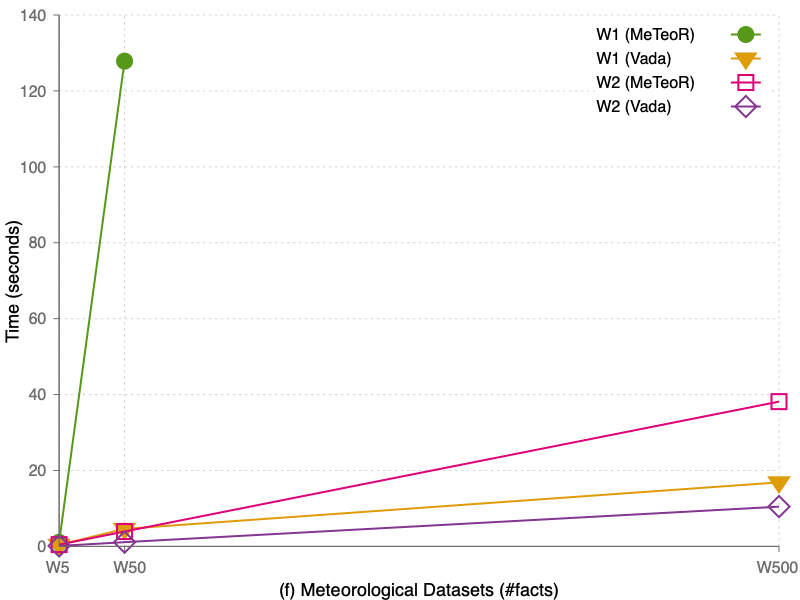}
    \caption{Larger version of the Meteorological experiments. }
    \label{fig:weather-graph}
\end{figure}

The results for the execution of programs W1 and W2 in Temporal Vadalog and MeTeoR (v. 1.0.15, mode \textit{seminaive}) are shown in Figure~\ref{fig:table-weather-results}. Figure~\ref{fig:weather-graph} is a larger version of the graph shown in the main contribution.

\begin{figure}[t]
%\begin{tabular}{llllllll}
\centering
\caption{The results of the Meteorological Experiments. The results are expressed in seconds. }
\label{fig:table-weather-results} {\tablefont\begin{tabularx}{\textwidth}{@{\extracolsep{\fill}}lllll}

\topline

Program:              & \multicolumn{2}{l}{\textbf{W1}}    & \multicolumn{2}{l}{\textbf{W2}}    \midline
\textbf{Dataset} & \textbf{Vadalog} & \textbf{MeTeoR} & \textbf{Vadalog} & \textbf{MeTeoR} \midline
\textbf{W5}   & 0.41             & 1.09            & 0.12             & 0.46            \\
\textbf{W50}  & 4.61             & 127.85          & 1.11             & 3.91            \\
\textbf{W500} & 16.88            & time out        & 10.49            & 38.14        
           
\botline
\end{tabularx}}
\end{figure}

\section{Time Series Experiments}

\subsection{Rules}

\subsubsection{EMA}

\begin{verbatim}
filterPredicate@[3,13596). % other ema values
startPredicate@[2,3]. % first ema value

b(X,Y) :- pad(X,Y), filterPredicate.
d(X,Y):- pad(X,Y).

emaStart(X,Z) :- <->[0,2] d(X,Y), Z=msum(Y).
emaStartInput(X,Z) :- emaStart(X,Y), startPredicate, Z = Y/3.
emaInput(X,Z,1) :- <->[1,1] ema(X,Y), Z=Y*(1-0.5).
emaInput(X,Z,2) :- b(X,Y), Z=Y*0.5.
ema(X,Z) :- emaInput(X,Y,V), Z=msum(Y,<V>).
emaMax(X,Z) :- ema(X,Y), Z=max(Y).
ema(X,Y):- emaStartInput(X,Y).
\end{verbatim}

\subsection{Results}

\begin{figure}
    \centering
    \includegraphics[width=0.95\textwidth]{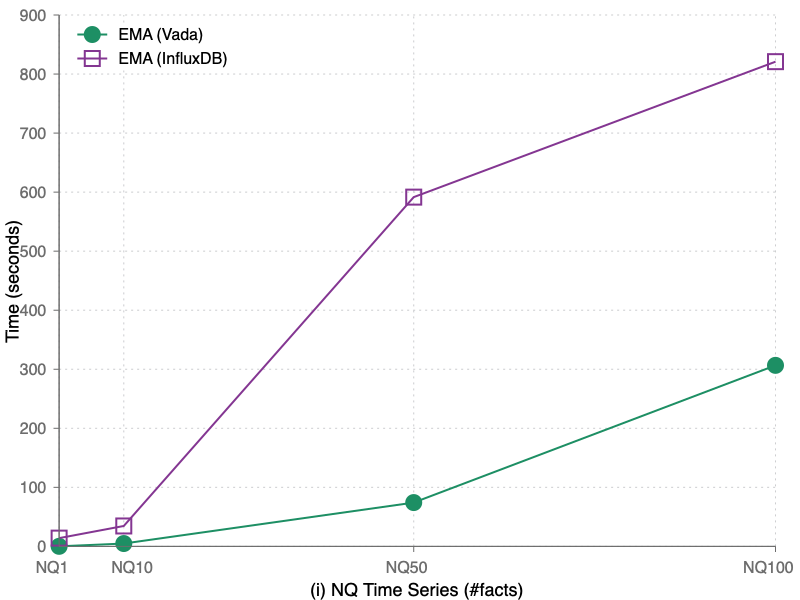}
    \caption{Larger version of the Time Series experiments on the Nasdaq Composite Index. }
    \label{fig:time-series-graph}
\end{figure}

The results for the execution of \textit{EMA (Exponential Moving Average)} in Temporal Vadalog and InfluxDB are shown in Figure~\ref{fig:table-timeseries-results}. Figure~\ref{fig:time-series-graph} is a larger version of the graph shown in the main contribution.

\begin{figure}[t]
%\begin{tabular}{llllllll}
\centering
\caption{The results of the Time Series Experiments. The results are expressed in seconds. }
\label{fig:table-timeseries-results} {\tablefont\begin{tabularx}{0.5\textwidth}{@{\extracolsep{\fill}}lll}

\topline
Program:         & \textbf{EMA} & \textbf{} \midline
\textbf{Dataset} & Vadalog      & InfluxDB  \midline
\textbf{NQ1}     & 0.0953       & 13.989    \\
\textbf{NQ10}    & 4.8127       & 34.57     \\
\textbf{NQ50}    & 74.1943      & 591.554   \\
\textbf{NQ100}   & 306.6180     & 820.992  
\botline
\end{tabularx}}
\end{figure}

\bibliographystyle{tlplike}
\bibliography{b}

%\printbibliography